\documentclass[11pt,twoside,a4paper,pdf]{article}  
\usepackage{graphicx,color}
\usepackage{times}
\usepackage{xspace}
\usepackage{booktabs}
\usepackage[english]{babel}
\usepackage{subfig}
\usepackage{amssymb}
\usepackage{cite}
\usepackage{xspace}
\usepackage{a4wide}
\usepackage{pdflscape}
\usepackage{amsmath,multicol,lineno,afterpage}
\usepackage[colorlinks=true, pdftex, citecolor=red, urlcolor=blue, ]{hyperref}

\newcommand{\mrm}[1]{\ensuremath{\mathrm{#1}}\xspace}

\newcommand{\tsc}[1]{\textsc{#1}}
\newcommand{\tit}[1]{\textit{#1}}

\newcommand{\EqRef}[1]{Eq.~(\ref{#1})\xspace}

\newcommand{\secRef}[1]{Section~\ref{#1}\xspace}
\newcommand{\secsRef}[1]{Sections~\ref{#1}\xspace}
\newcommand{\TabRef}[1]{Table~\ref{#1}\xspace}
\newcommand{\TabsRef}[1]{Tables~\ref{#1}\xspace}
\newcommand{\FigRef}[1]{Fig.~\ref{#1}\xspace}
\newcommand{\FigsRef}[1]{Figs.~\ref{#1}\xspace}
\newcommand{\FigureRef}[1]{Figure~\ref{#1}\xspace}

\newcommand{\inst}[1]{$^{#1}$}
\renewcommand{\and}{, }

\newcommand{\hpp}{\tsc{Herwig++}\xspace}
\newcommand{\vc}{\tsc{Vincia}\xspace}
\newcommand{\Hpp}[1]{\tsc{Herwig++~{#1}}}
\newcommand{\Hw}[1]{\tsc{Herwig~{#1}}}
\newcommand{\Py}[1]{\tsc{Pythia~{#1}}}
\newcommand{\Vc}[1]{\tsc{Vincia~{#1}}}
\newcommand{\Ari}[1]{\tsc{Ariadne~{#1}}}

\newcommand{\roo}{\tsc{Root}\xspace}
\newcommand{\roou}{\tsc{RooUnfold}\xspace}

\newcommand{\epem}{\ensuremath{e^+e^-}}
\newcommand{\qqbar}{\ensuremath{q\bar q}}

\graphicspath{{images/}}

\begin{document}

\vspace*{-1.8cm}\begin{minipage}{\textwidth}
\flushright
OPAL PR435 \\
KA-TP-09-2015 \\
COEPP-MN-15-2 \\
MPP-2015-98
\end{minipage}
\vskip1.25cm
{\Large\bf
\begin{center}
Measurement of observables sensitive to coherence effects in
hadronic $Z$ decays with the OPAL detector at LEP
\end{center}}
\vskip5mm
{\begin{center}
{\large 
N.~Fischer\inst{1,2}\and 
S.~Gieseke\inst{1}\and
S.~Kluth\inst{3}\and
S.~Pl\"atzer\inst{4,5}\and
P.~Skands\inst{2,6}\and and \\
The OPAL collaboration\cite{opal}
}
\end{center}
\vskip3mm
\begin{center}
\parbox{0.9\textwidth}{
\inst{1}: Institute for Theoretical Physics, Karlsruhe Institute of Technology, Karlsruhe, Germany\\
\inst{2}: School of Physics and Astronomy, Monash University, Melbourne, Australia\\
\inst{3}: Max-Planck-Institute for Physics, Munich, Germany\\
\inst{4}: Institute for Particle Physics Phenomenology, Durham University, Durham, United Kingdom\\
\inst{5}: School of Physics and Astronomy, University of Manchester, Manchester, United Kingdom\\
\inst{6}: Theoretical Physics, CERN, Geneva, Switzerland\\
}
\end{center}
\vskip5mm
\begin{center}
\parbox{0.85\textwidth}{
\begin{center}
\textbf{Abstract}
\end{center}\small
A study of QCD coherence is presented based on a sample of about
397~000 \epem\ hadronic annihilation events collected at
$\sqrt{s}=91$~GeV with the OPAL detector at LEP.  The study is based
on four recently proposed observables that are sensitive to coherence
effects in the perturbative regime.  The measurement of these
observables is presented, along with a comparison with the predictions
of different parton shower models. The models include both
conventional parton shower models and dipole antenna models.
Different ordering variables are used to investigate their influence
on the predictions.  }
\end{center}
\vspace*{1cm}

\section{Introduction \label{sec:intro}}

Processes involving the strong interaction, described in the standard
model (SM) by quantum chromodynamics (QCD), dominate in high energy
particle collisions.  It is therefore important to account for QCD
effects and to model them accurately.  Colour coherence, the
destructive interference effect between colour-connected partons, is
an important aspect of high energy collisions and QCD parton cascades.
Coherence is itself a subject of considerable interest, and QCD offers
a situation in which coherence effects in a perturbative framework can
be studied in a uniquely precise way.  Furthermore, by testing
different theoretical schemes for coherence, QCD Monte Carlo (MC)
event generators (see
Refs.~\cite{Buckley:2011ms,Beringer:1900zz,Seymour:2013ega,Gieseke:2013eva}
for recent reviews) can be modified to better describe the results of
experiments.  For example, in new-physics searches at the CERN LHC,
QCD multijet events often represent the most difficult SM background
to characterize. Improvements in the reliability of QCD event
generators may help to better constrain this background.

The \epem\ annihilation process offers a favorable environment to
study colour coherence, because the lack of strong interactions in the
initial state allows simple and conclusive comparisons between
experiment and theory. Previous studies of coherence in
\epem\ annihilation events are presented, for example, in
Refs.~\cite{Abbiendi:2006qr,Abdallah:2004uu}.  Within the context of a
QCD shower, coherence implies an ordering condition, such as a
requirement that each subsequent emission angle in the shower be
smaller than the previous
angle~\cite{Mueller:1981ex,Dokshitzer:1982fh}.  However, there are
many ambiguities in the definition of the ordering variable and in its
implementation. In this study, we present the first experimental tests
of recently proposed~\cite{Fischer:2014bja} observables designed to
discriminate between coherence schemes.  The data were collected with
the OPAL detector at the CERN LEP collider at a centre-of-mass energy
of $\sqrt{s}=91$~GeV. The observables examined here are based on
four-jet \epem\ annihilation configurations in which a soft gluon is
emitted in the context of a three-jet topology, with two of the three
jets approximately collinear. This event configuration has been shown
to be favorable for the manifestation of
coherence~\cite{Platzer:2009jq} and sensitive to the choice of the
ordering variable in the shower~\cite{AlcarazMaestre:2012vp}.

We examine six different models for coherence, which are implemented
in currently available QCD MC event generator programs. Specifically,
we compare the default $\tilde q^2$ parton shower of
\hpp~\cite{Gieseke:2003rz} with angular-ordering, the
$p_{\perp\mrm{dip}}^2$- and $q_\mrm{dip}^2$-ordered dipole showers of
\hpp~\cite{Platzer:2009jq}, the default
$p_{\perp\mrm{evol}}^2$-ordered shower of
\Py{8}~\cite{Sjostrand:2004ef}, and the $p_{\perp\mrm{ant}}^2$- and
$m_{\mrm{ant}}^2$-ordered showers of \tsc{Vincia}~\cite{Giele:2007di},
a plugin to the \Py{8} event generator that replaces the \Py{8} shower
with a shower model based on antenna functions.  The definitions of
the ordering variables are given in the next section, with more
details presented in Ref.~\cite{Fischer:2014bja}.

This paper is organized as follows. In \secRef{sec:theory}, we define
the observables to be used in the
analysis. \secsRef{sec:experiment},~\ref{sec:samples} and
\ref{sec:analysis} present the detector, the data sample and
simulation, and the data analysis, respectively.  The results are
presented in \secRef{sec:comp} and our conclusions in
\secRef{sec:conclusions}.

\section{Theoretical concepts \label{sec:theory}}

\subsection{Observables \label{sec:observables}}

We consider hadronic events from \epem\ annihilation at the $Z$ boson
peak and use the Durham $k_T$ clustering
algorithm~\cite{Catani:1991hj} to cluster all particles of an event
into jets, keeping track of the clustering scales along the way.  The
algorithm begins by assigning all particles in an event to a list.
Each entry in the list is called a jet.  The algorithm then computes,
for all pairs of four-momenta $i$ and $j$ in the event, the distance
measure
\begin{align}
y_{ij}=2\mrm{min}(E_i^2,E_j^2)(1-\cos\theta_{ij})/s~,
\end{align}
where $E_i$ and $E_j$ are the corresponding energies and $\theta_{ij}$
is the angle between objects $i$ and $j$. The center-of-mass
energy-squared is denoted by $s$.  The pair of objects with the
smallest $y_{ij}$ is combined by summing their four-momenta and the
sum is added to the list while the original four-momenta are
removed. This procedure is iterated until one entry is left in the
list. To obtain an inclusive four-jet event sample in the perturbative
regime we impose an explicit requirement on the value of the
clustering scale at which the event goes from having four to having
three jets. Denoting this scale (given by the value of
$\mrm{min}(y_{ij})$ evaluated at the stage when the event has been
clustered to four jets) by $y^{4\to3}$, we require $y^{4\to3}>0.0045$
(corresponding to $\ln(y^{4\to3})~>~-5.4$), as in
Ref.~\cite{Fischer:2014bja}. This value originates from a compromise;
on the one hand a smaller value results in a data sample with greater
statistical precision, and on the other hand a larger value provides a
more direct representation of the shower properties.

We investigate four different observables, where for the first three
we consider the event clustered into four jets, and order the jets in
energy. To be sensitive to coherence, the angles between the jets are
constrained such that the first (hardest) jet lies back-to-back to a
nearly collinear jet pair, formed by the second and third jet:
$\theta_{12}>2\pi/3$, $\theta_{13}>2\pi/3$, and $\theta_{23}<\pi/6$.
The event topology resulting from these requirements is shown in
\FigRef{fig:sketches}~\tit{a)}.  To investigate QCD colour coherence
effects we examine the following observables:

\begin{figure}[t]
\tit{a)}\hspace*{8cm}\tit{b)} \vspace*{-3mm} \\
\includegraphics[trim=1cm 25cm 14.5cm 1cm,clip,width=7.5cm]{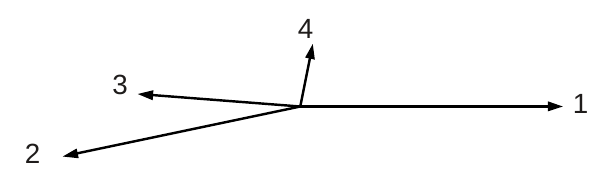}
\hspace*{5mm}
\includegraphics[trim=1cm 25cm 14.5cm 1cm,clip,width=7.5cm]{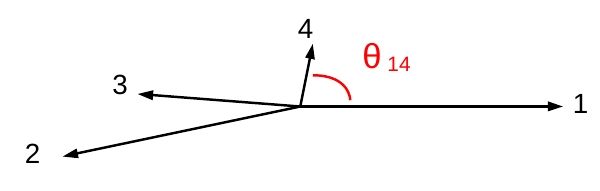} \vspace*{3mm} \\
\tit{c)}\hspace*{8cm}\tit{d)} \vspace*{-3mm} \\
\includegraphics[trim=1cm 25cm 14.5cm 1cm,clip,width=7.5cm]{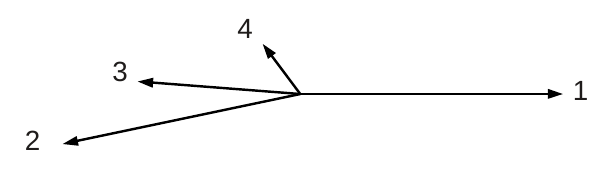}
\hspace*{5mm}
\includegraphics[trim=1cm 25cm 14.5cm 1cm,clip,width=7.5cm]{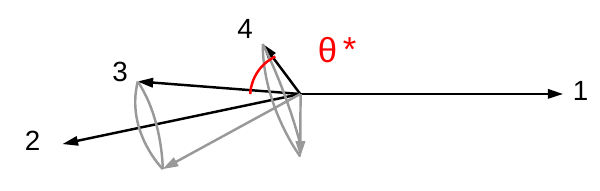} \vspace*{3mm} \\
\tit{e)}\hspace*{8cm}\tit{f)} \vspace*{-10mm} \\
\includegraphics[trim=1cm 19.25cm 15cm 7cm,clip,width=7cm]{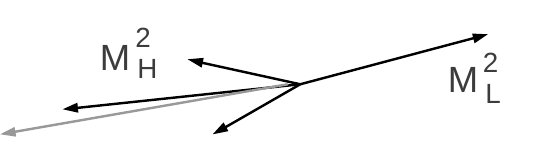}
\hspace*{10mm}
\includegraphics[trim=1cm 16.75cm 15cm 8.5cm,clip,width=7cm]{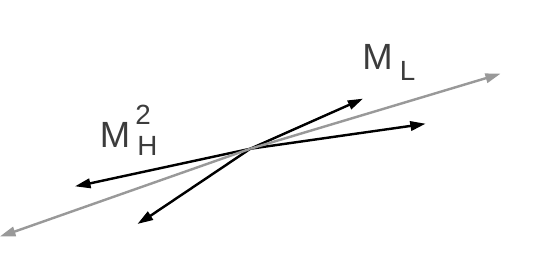}
\caption{ \tit{a)} The event topologies resulting from four-jet events
  with requirements on the angles between the jets:
  $\theta_{12}>2\pi/3$, $\theta_{13}>2\pi/3$, and
  $\theta_{23}<\pi/6$. \tit{b)} Illustration of the observable
  $\theta_{14}$, the angle between the first and fourth jet in the
  latter events.  \tit{c)} The event topologies resulting from
  four-jet events with requirements $\theta_{12}>2\pi/3$,
  $\theta_{13}>2\pi/3$, $\theta_{23}<\pi/6$, and
  $\theta_{24}<\pi/2$. \tit{d)} Illustration of the observable
  $\theta^*=\theta_{24}-\theta_{23}$, the difference in opening
  angles.  \tit{e)} The sketch shows event topologies where the third
  and fourth clusterings occur within the same jet and hence the mass
  ratio $M_L^2/M_H^2$ is small. \tit{f)} Events with large mass
  ratios, where the third and fourth clusterings occur on opposite
  sides of the event.  }
\label{fig:sketches}
\end{figure}

\begin{itemize}
\item $\theta_{14}$, proposed in Ref.~\cite{AlcarazMaestre:2012vp}:
  \\ The emission angle of the soft fourth jet with respect to the
  first jet; a sketch can be found in \FigRef{fig:sketches}~\tit{b)}.
\item $\theta^*$, proposed in Ref.~\cite{Platzer:2009jq}: \\ A
  restriction on the angle between the second and fourth jet,
  $\theta_{24}<\pi/2$, is imposed in order to require the fourth jet
  to be close in angle to the nearly collinear (23) jet pair, see
  \FigRef{fig:sketches}~\tit{c)}. The observable is the difference in
  opening angles, $\theta^*=\theta_{24}-\theta_{23}$, and is sensitive
  to coherent emission from the (23) jet system. A sketch of this
  observable is shown in \FigRef{fig:sketches}~\tit{d)}.
\item $C_2^{(1/5)}$, proposed in Ref.~\cite{Larkoski:2013eya}: \\ In
  general we have the freedom to chose the exponent $\beta$ of the
  2-point energy correlation double ratio $C_2^{(\beta)}$. With the
  choice $\beta=1/5$ and a nearly collinear (23) jet pair we can
  express the 2-point double ratio as $C_2^{(1/5)} \approx
  E_4\theta_{14}^{1/5}\theta_{23\,4}^{1/5}E_\text{vis}/(E_1E_{23}\theta_{1\,23}^{1/5})$,
  with the total visible energy $E_\text{vis}$ in the event.
  $\theta_{23\,4}$ denotes the angle between the softest jet and the
  (23) jet pair and analogously for $\theta_{1\,23}$.  Therefore the
  2-point double ratio is mainly sensitive to the relative energy of
  the fourth jet.
\end{itemize}

Strong ordering in the parton shower refers to strong ordering of the
clustering scales, $y^{3\to2}\gg y^{4\to3} \gg \ldots$, with
$y^{(n+1)\to n}$ the value of the jet distance parameter in the Durham
algorithm for which the configuration changes from $n+1$ to $n$ jets.
In contrast, events with, e.g., $y^{4\to3}\sim y^{3\to2}$ are more
sensitive to the ordering condition and to situations where the same
parton participates in two splitting processes, hence to effective
$1\to 3$ splittings. For the final observable, we cluster events into
two jets and apply the restriction $y^{4\to3} >0.5\, y^{3\to2}$. This
forces events into a compressed hierarchy, i.e., a hierarchy without
strong ordering.  The investigated observable is:
\begin{itemize}
\item $\rho=M_L^2/M_H^2$, proposed in
  Ref.~\cite{AlcarazMaestre:2012vp}: \\ The ratio of the invariant
  masses-squared of the jets at the end of the clustering process,
  ordered such that $M_L^2 \le M^2_H$. For ``same-side'' events, where
  one $1\to 3$ splitting occurs, the mass ratio is close or equal to
  zero, whereas for ``opposite-side'' events with $1\to 2 \otimes 1\to
  2$ splittings, the mass ratio is larger. In
  \FigRef{fig:sketches}~\tit{e)} and \tit{f)} we illustrate examples
  of these event topologies. For references to heavy jet masses see,
  e.g. Refs.~\cite{Chandramohan:1980ry,Clavelli:1981yh}.
\end{itemize}

To exhibit the differences between the theory models more clearly, we 
introduce the asymmetry for a given observable $x$,
\begin{align}\label{eq:asymmetry}
\frac{N_\text{left}}{N_\text{right}} = \dfrac{\sum\limits_{i~\text{\tiny with}~x(i)<x_0}n_i}
{\sum\limits_{i~\text{\tiny with}~x(i)>x_0}n_i} ~,
\end{align}
where $n_i$ is the number of events in histogram bin $i$ and $x(i)$ is
the bin center. The dividing point $x_0$ separates the regions with
small and large values of $x$. We use this asymmetry for three of the
four observables, $\theta^*$, $C_2^{(1/5)}$, and $\rho$, and thus
introduce three dividing points: $\theta^*_0$, $C_{2,0}^{(1/5)}$, and
$\rho_0$.

As in Ref.~\cite{Fischer:2014bja}, we divide the full $\theta_{14}$
range into three regions labelled ``towards'' (small $\theta_{14}$),
``central'' (intermediate $\theta_{14}$), and ``away'' (large
$\theta_{14}$), denoted ``T'', ``C'', and ``A'' respectively. 
In the towards region, the first and fourth jets are collinear, while
they are back-to-back in the away region. Events in which the fourth
jet represents a wide-angle emission from the three-jet system
populate the central region. We then
consider the ratio between regions $R_j$ and $R_k$,
\begin{align}\label{eq:A14ratios}
\frac{R_j}{R_k} = \dfrac{\sum\limits_{i\subset R_j}n_i}
{\sum\limits_{i\subset R_k}n_i}~.
\end{align}
We define 9 different versions of the ratio, with different definitions 
of the regions, which are given in \TabRef{tab:DefRegionsA14}.

\begin{table}[t]
\begin{center}
\begin{tabular}{ccccc} 
 & & Central/Towards & Central/Away & Towards/Away \vspace*{2mm} \\
\toprule
\# & Central region & Towards region & Away region & Towards region \\ 
\midrule
1 & $0.4<\theta_{14}/\pi<0.6$ & $\theta_{14}/\pi<0.3$ & $\theta_{14}/\pi>0.6$ & $\theta_{14}/\pi<0.3$ \\
2 & $0.4<\theta_{14}/\pi<0.6$ & $\theta_{14}/\pi<0.2$ & $\theta_{14}/\pi>0.7$ & $\theta_{14}/\pi<0.3$ \\
3 & $0.4<\theta_{14}/\pi<0.6$ & $\theta_{14}/\pi<0.4$ & $\theta_{14}/\pi>0.8$ & $\theta_{14}/\pi<0.3$ \\
4 & $0.45<\theta_{14}/\pi<0.55$ & $\theta_{14}/\pi<0.3$ & $\theta_{14}/\pi>0.6$ & $\theta_{14}/\pi<0.2$ \\
5 & $0.45<\theta_{14}/\pi<0.55$ & $\theta_{14}/\pi<0.2$ & $\theta_{14}/\pi>0.7$ & $\theta_{14}/\pi<0.2$ \\
6 & $0.45<\theta_{14}/\pi<0.55$ & $\theta_{14}/\pi<0.4$ & $\theta_{14}/\pi>0.8$ & $\theta_{14}/\pi<0.2$ \\
7 & $0.35<\theta_{14}/\pi<0.65$ & $\theta_{14}/\pi<0.3$ & $\theta_{14}/\pi>0.6$ & $\theta_{14}/\pi<0.4$ \\
8 & $0.35<\theta_{14}/\pi<0.65$ & $\theta_{14}/\pi<0.2$ & $\theta_{14}/\pi>0.7$ & $\theta_{14}/\pi<0.4$ \\
9 & $0.35<\theta_{14}/\pi<0.65$ & $\theta_{14}/\pi<0.4$ & $\theta_{14}/\pi>0.8$ & $\theta_{14}/\pi<0.4$ \\
\bottomrule
\end{tabular}
\end{center}
\caption{Definitions of the $\theta_{14}$ intervals used for the
  asymmetry ratios defined in \EqRef{eq:A14ratios}. We define 9 different 
  versions of the ratio with the labeling of the regions given in the
  first column. The ratio between the results in
  the central and towards regions is based on the definitions in
  columns two and three, between the central and away regions on the
  definitions in columns two and four, and between the towards and
  away regions on the definitions in columns four and five. Taken from
  Ref.~\cite{Fischer:2014bja}.}
\label{tab:DefRegionsA14}
\end{table}

\subsection{Theory models \label{sec:models}}
For parton showers based on $1\to2$ splittings of a parton $I$ to
daughters $i$ and $j$, momentum conservation requires that the
virtuality of the branching parton must be compensated for by a recoil
somewhere else in the event; we refer to parton $I$ as the ``emitter''
and to the parton (system) absorbing the recoil as the ``recoiler''.

The six different theory models for the parton shower, mentioned in
\secRef{sec:intro}, are based on different formalisms and radiation functions:
\begin{itemize}
\item In the collinear DGLAP
  formalism~\cite{Gribov:1972ri,Altarelli:1977zs,Dokshitzer:1977sg},
  each parton is evolved separately and undergoes $1\to2$ like
  branchings, which we denote $p_I\to p_ip_j$. In order to respect QCD
  coherence properties, a specific choice for the evolution
  variable~\cite{Gieseke:2003rz,Marchesini:1983bm}, or additional
  vetos~\cite{Bengtsson:1986et}, are applied. The momentum balancing
  can either include all partons of the event, which we refer to as
  global recoils, or only one recoiler parton, which we refer to as
  local recoil.
\item Another formalism is based on the Catani-Seymour (CS) dipole
  functions~\cite{Catani:1996vz}, where a single parton emission from
  a pair of partons is considered. We denote the momenta involved in
  this splitting process with $p_Ip_K\to p_ip_jp_k$. The full
  splitting probability is partitioned into two pieces, corresponding
  to partons $I$ and $K$, respectively, acting as the emitter with the
  other acting as the recoiler. The recoil is limited to the
  longitudinal direction of the recoiler parton in the rest frame of
  $I$ and $K$.  If the dipole shower uses an evolution with ordering
  in transverse momentum, the shower correctly reproduces the soft
  properties of QCD.
\item In the QCD antenna (also called Lund
  dipoles)~\cite{Gustafson:1987rq,Kosower:2003bh} picture, there is no
  fundamental distinction between the emitter and the recoiler. Each
  colour-connected parton pair of an event is represented by an
  antenna and undergoes a splitting process of the form $p_Ip_K\to
  p_ip_jp_k$ with a $2\to 3$ recoil prescription.  A single antenna
  thereby accounts for the equivalent of two CS dipoles.
\end{itemize}
The theory models we investigate here span all the above
formalisms. For the ordering variables we use the notation
$Q_I^2=(p_i+p_j)^2$, $Q_K^2 =(p_j+p_k)^2$, and
$M_{IK}^2=(p_I+p_K)^2=(p_i+p_j+p_k)^2$, for the splitting processes as
stated above.  For the DGLAP-based models, the parton $K$ acts as the
recoiler and can either represent a single parton (\Py{8}) or multiple
partons (\hpp).

In the following we briefly describe the main differences between the
theory models used in this paper, mostly concentrated on the aspects
described above.  \hpp $\tilde q^2$~\cite{Gieseke:2003rz}, a parton
shower model based on DGLAP splitting kernels, uses global
recoils. The evolution is ordered in a variable proportional to energy
times angle,
\begin{align}
\tilde q^2 &= \frac{Q_I^2M_{IK}^4}{Q_K^2(M_{IK}^2-Q_I^2-Q_K^2)}~.
\end{align}
The shower includes a matrix-element correction for the first emission
and uses two-loop running of $\alpha_s$. The QCD coherence properties
are respected due to the angular ordering of the parton branching
cascade.  The second shower model in the \hpp event generator is \hpp
$p_{\perp\mrm{dip}}^2$~\cite{Platzer:2009jq}, which is based on
partitioned CS dipoles with local recoils within dipoles. The ordering
variable is the relative transverse momentum of the splitting pair,
\begin{align}
p^2_{\perp\mrm{dip}} &= \frac{Q_I^2Q_K^2(M_{IK}^2-Q_I^2-Q_K^2)}{(M_{IK}^2-Q_I^2)^2}~.
\end{align}
We do not apply matching or matrix-element corrections and use
one-loop running of $\alpha_s$.  The dipole shower with ordering in
transverse momentum respects QCD coherence.  As an alternative we use
the same shower model, but with a different ordering variable. \hpp
$q_\mrm{dip}^2$~\cite{Platzer:2009jq} orders the shower cascade in
virtuality of the splitting pair,
\begin{align}
q_\mrm{dip}^2 &= Q_I^2~.
\end{align}
As before we do not apply matching or matrix-element corrections and
use one-loop running of $\alpha_s$.  \vc
$p_{\perp\mrm{ant}}^2$~\cite{Giele:2007di} is a shower model based on
antenna functions with local recoils within antennae. The ordering
variable is the transverse momentum of the antenna,
\begin{align}
p_{\perp\mrm{ant}}^2 &= \frac{Q^2_I Q^2_K}{M_{IK}^2}~.
\end{align}
Matrix-element corrections at LO~\cite{Giele:2011cb} and
NLO~\cite{Hartgring:2013jma} are switched off and we use one-loop
running of $\alpha_s$. Colour coherence is respected, since it is an
intrinsic property of the antenna functions. Transverse momentum as
the evolution variable is the preferred choice in \vc, as has been
shown in Ref.~\cite{Hartgring:2013jma}.  However, we also use \vc
$m_\mrm{ant}^2$~\cite{Giele:2007di} as an alternative to the
transverse momentum ordering, which orders the shower evolution in
antenna mass, defined as
\begin{align}
m_{\mrm{ant}}^2 &= \min(Q^2_I,Q^2_K)~.
\end{align}
The last shower model is \Py{8} $p_{\perp\mrm{evol}}^2$~\cite{Sjostrand:2004ef}, a parton
shower based on DGLAP splitting kernels and ordered in transverse momentum, defined as
\begin{align}
p^2_{\perp\mrm{evol}} &= \frac{Q_I^2(M^2_{IK}-Q^2_K)(Q^2_I+Q^2_K)}{(M^2_{IK}+Q^2_I)^2}~.
\end{align}
In contrast to the angular ordered \hpp shower, local recoils within
dipoles are applied. A matrix-element correction for the first
emission is included and we use one-loop running of $\alpha_s$. To
obtain QCD coherence properties, the shower applies angular vetoes.

Besides the shower models used in this paper, there are several other
models: \Ari~\cite{Lonnblad:1992tz}, based on antenna functions, which
is very similar to \vc; the CS dipole shower models of Weinzierl et
al.~\cite{Dinsdale:2007mf}, and \tsc{Sherpa}~\cite{Schumann:2007mg},
which are similar to the \hpp dipole shower; the deductor by Nagy and
Soper~\cite{Nagy:2014mqa}, which is not interfaced with a
hadronization model; and the virtuality-ordered final-state showers of
\Py~\cite{Bengtsson:1986et,Bengtsson:1986hr},
\tsc{Nlljet}~\cite{Kato:1990as} and \tsc{Herwiri}~\cite{Ward:2010af}.

To compare the models on as equal a footing as possible, the shower
and hadronization parameters have been readjusted with the
\tsc{Professor}~\cite{Buckley:2009bj} tuning system, utilizing LEP
data available through \tsc{Rivet}~\cite{Buckley:2010ar}.  The tuning
procedure is described in Ref.~\cite{Fischer:2014bja}, where the
resulting parameter values can also be found.

\section{OPAL experiment \label{sec:experiment}}

The OPAL experiment at LEP operated between August 1989 and November
2000.  The detector components were arranged around the beam pipe, in
a layered structure. A detailed description can be found in
Refs.~\cite{Ahmet:1990eg,Anderson:1997xwa,Aguillion:1998pz}.  The
tracking system consisted of a silicon microvertex detector, an inner
vertex chamber, a jet chamber, and chambers outside the jet chambers
to improve the precision of the $z$-coordinate\footnote{OPAL uses the
  right-handed coordinate system defined with the $x$-axis pointing
  towards the center of the LEP ring, the positive $z$ points along
  the direction of the $e^-$ beam and the $y$-axis upwards.  $r$ is
  the coordinate normal to the beam axis and the polar angle $\theta$
  and the azimuthal angle $\varphi$ are defined with respect to $x$
  and $z$.} measurement. The jet chamber was approximately $4~\mrm
m$ long and had an outer radius of about $1.85~\mrm m$. This device
had $24$ sectors each containing $159$ sense wires spaced by 1 cm.
All tracking systems were located inside a solenoidal magnet, which
provided a uniform axial magnetic field of $0.435~\mrm{T}$ along the
beam axis.  The magnet was surrounded by a lead glass electromagnetic
calorimeter and a sampling hadron calorimeter.  The electromagnetic
calorimeter consisted of $11704$ lead glass blocks, divided into
barrel and endcap sections, covering $98\%$ of the solid angle.
Outside the hadron calorimeter, the detector was surrounded by a
system of muon chambers.  Similar layers of instrumentation were
located in the endcap regions.

Since the energy resolution of the electromagnetic calorimeter is
better then that of the hadron calorimeter, the resolution of jet
directions and energies is not significantly improved by incorporating
hadron calorimeter information. Thus, our analysis relies exclusively
on charged particle information recorded in the tracking detectors and
on clusters of energy deposited in the electromagnetic calorimeter.

\section{Data and MC samples \label{sec:samples}}

In the first phase of LEP operation, denoted LEP1 (1989 to 1995), the
\epem\ center-of-mass energy was chosen to lie at or near the mass of
the $Z$ boson, $\sqrt{s}\approx 91~\mrm{GeV}$.  During the second
phase of operation, denoted LEP2 (1995-2000), the center-of-mass
energy was increased in successive steps from $130$ to
$209~\mrm{GeV}$.  Interspersed at various times during the LEP2
operation, calibration runs were collected at the $Z$ boson peak.  In
this analysis, we utilize data collected at $\sqrt{s}=91.2$~GeV during
the the LEP2 calibration runs. This allows us to exploit conditions
when the detector was operating in its final, most advanced
configuration. In addition, this will facilitate possible future
comparisons with data collected under essentially identical conditions
at higher energies. We use a sample corresponding to an integrated
luminosity of $14.7~\mrm{pb}^{-1}$.  This sample is of sufficient size
that systematic uncertainties dominate the statistical terms.  To
correct the data in order to account for experimental acceptance and
efficiency, simulated event samples produced with MC event generators
are used. The process $\epem\to\qqbar$ is simulated using
\Py{6.1}~\cite{Sjostrand:2000wi} at $\sqrt{s}=91.2$~GeV.
Corresponding samples using
\Hw{6.2}~\cite{Marchesini:1991ch,Corcella:2000bw} are used for
systematic checks.  We examine the MC events at two levels. We refer
to ``hadron level'' as events without event selection, and without
simulation of the detector acceptance and resolution, for which all
particles with lifetimes less than $300~\mrm{ps}$ decay. In contrast,
``detector level'' refers to MC events that are processed through the
\tsc{Geant4}-based~\cite{Agostinelli:2002hh} simulation of the OPAL
detector, called \tsc{Gopal}~\cite{Allison:1991bf}, and that have been
reconstructed using the same software procedures that are applied to
the data.  The MC events generated for the detector-level samples are
the same as the hadron-level samples except that $K^0_S$ mesons and
weakly decaying hyperons are declared to be stable, as these particles
can interact with detector material before decaying, and so their
decays are handled within the \tsc{Geant} framework.
  
In addition, for comparisons with the corrected data, large samples 
of hadron-level MC events are employed, using the event 
generators \Hpp{2.7.0}~\cite{Bellm:2013lba}, \Py{8.176}~\cite{Sjostrand:2007gs}, 
and \Vc{1.1.0}~\cite{Giele:2007di} interfaced with the hadronization model 
of \Py{8.176}.

\section{Data analysis \label{sec:analysis}}

\subsection{Selection of events}

The same criteria for the selection of charged tracks and
electromagnetic clusters are applied as described in
Ref.~\cite{Abbiendi:2004qz}. Charged tracks are required to have
transverse momentum relative to the beam axis larger than
$0.15~\mrm{GeV}$, and photons to have energies larger than
$0.10~\mrm{GeV}$ ($0.25~\mrm{GeV}$) in the barrel (endcap) region of
the electromagnetic calorimeter. The selection of hadronic
annihilation events is the same as described in
Ref.~\cite{Alexander:1991qw}. Briefly, a minimum of five charged
tracks is required, and a containment condition
$|\cos\theta_{\mrm{T}}|<0.90$ is applied, where $\theta_{\rm{T}}$ is
the polar angle of the thrust axis~\cite{Brandt:1964sa,Farhi:1977sg}
with respect to the beam axis, calculated using all accepted charged
tracks and electromagnetic clusters. A total of $397\,452$ candidate
hadronic annihilation events are selected, with a negligible expected
background.

Since the energy loss due to initial-state radiation is highly
suppressed at the $Z$ peak, we do not apply a cut to that effect.
However, radiative corrections are applied by requiring
$\sqrt{s}-\sqrt{s'}<1~\mrm{GeV}$ for the MC detector-level samples
used to correct the data, where $\sqrt{s'}$ is the effective
center-of-mass energy after initial-state radiation.

\subsection{Reconstruction and correction}

For each of the accepted events, the values of all observables
described in \secRef{sec:intro} are computed.  To avoid
double-counting of energy between tracks and electromagnetic clusters,
an energy-flow algorithm~\cite{Ackerstaff:1997nga,Abbiendi:1999sy} is
applied, which matches the tracks and clusters and retains only those
clusters that are not associated with a track.

\begin{figure}[p]
\includegraphics[width=\textwidth]{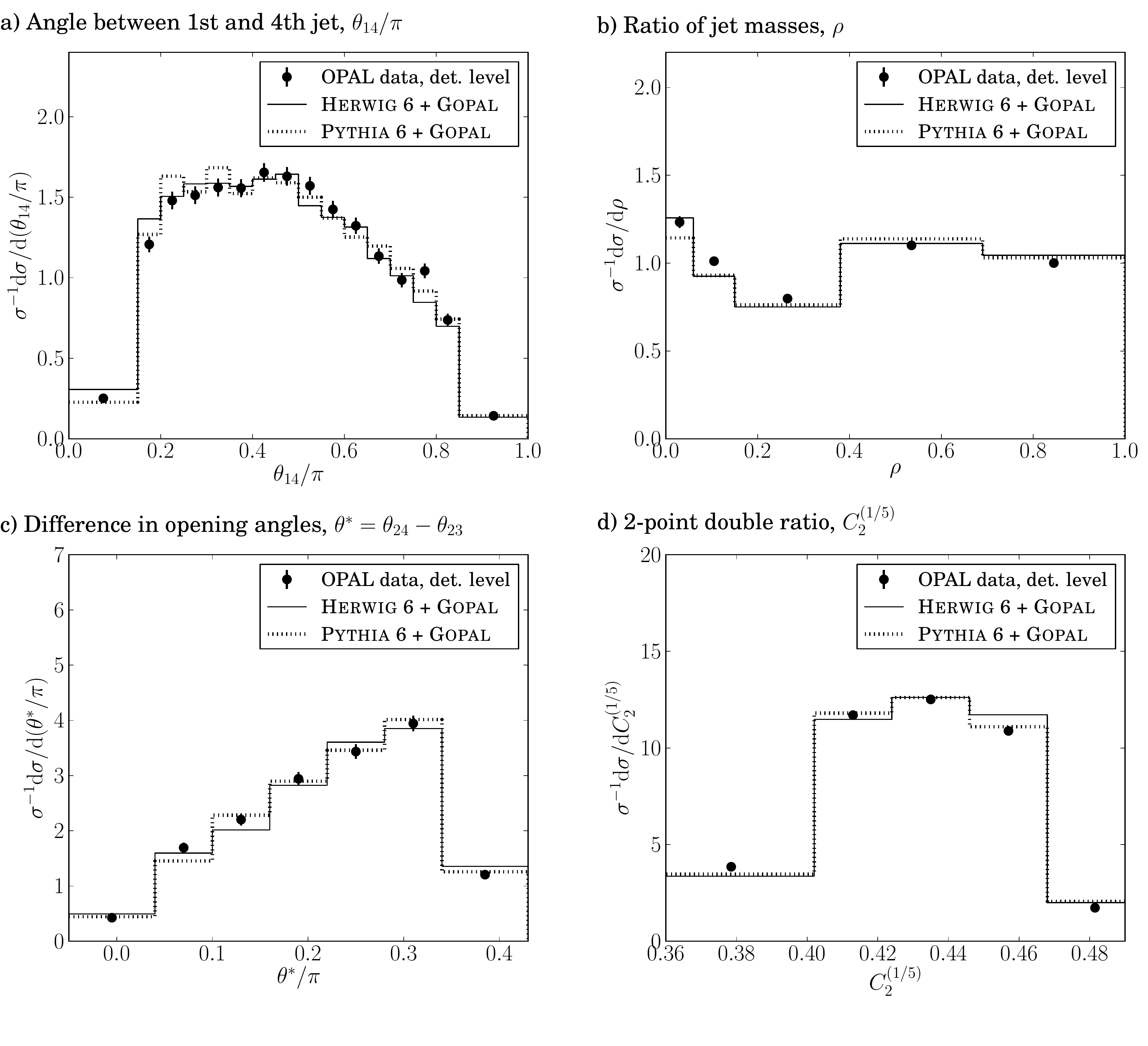}
\caption{The uncorrected distributions of \tit{a)} the emission angle
  $\theta_{14}$, \tit{b)} the mass ratio $\rho=M_L^2/M_H^2$, \tit{c)}
  the difference in opening angles $\theta^*$, and \tit{d)} the
  2-point double ratio $C_2^{(1/5)}$, in comparison with the
  predictions of the \Hw{6}\ and \Py{6}\ Monte Carlo event generators
  at the detector level. The error bars indicate the statistical
  uncertainties.}
\label{fig:detlevel}
\end{figure}

\FigureRef{fig:detlevel} shows a comparison of the uncorrected data
with the detector-level predictions of \Py{6}\ and \Hw{6}\ for the
$\theta_{14}$, $\rho$, $\theta^*$, and $C_2^{(1/5)}$ variables. The
$\theta_{14}$ and $\theta^*$ variables are normalized by a factor
of~$\pi$.  The simulations are seen to provide a generally adequate
description of the measurements.

To correct the data for detector and resolution effects, we implement
an unfolding procedure based on the \roou~\cite{Adye:2011gm}
framework. We use the iterative Bayes method, as proposed by
D'Agostini~\cite{D'Agostini:1994zf}, with four iterations, which is
the recommendation from Ref.~\cite{Adye:2011gm}.  A necessary
ingredient for the unfolding is the response matrix of the MC event
generator used for the correction procedure. For the standard
analysis, \Py{6}\ is used to determine the response matrix. The
response matrix gives the bin-to-bin migration from the hadron to the
detector level, and vice versa.  In order to obtain reliable results
for the corrected distributions, we adjust the bin widths of the
histograms such that the probability for a hadron-level event to
migrate to a different bin at the detector level is less than $50\%$.

The corrected distributions are presented in \FigRef{fig:data} and
\TabsRef{tab:dataBinNorm1}-\ref{tab:dataBinNorm2}. 
\TabsRef{tab:dataBinNorm1}-\ref{tab:dataBinNorm2}
include the covariance matrices calculated with \roou. The statistical
uncertainties are given by the square root of the corresponding
diagonal element in the covariance matrices. Systematic uncertainties
are discussed in \secRef{sec:systematics}.  \FigureRef{fig:data}
includes the predictions of \Py{6}\ and \Hw{6}\ at the hadron level.
The differences between the MC predictions and the data are seen to be
similar to those observed at the detector level
(\FigRef{fig:detlevel}), demonstrating that the correction procedure
does not introduce a discernible bias.

The values of the derived distributions, i.e., the ratios of the
different regions for $\theta_{14}$ and the asymmetry for the other
observables, are listed in \TabRef{tab:dataAS}. The quantities are
determined by summing and dividing the histogram entries. The
statistical uncertainties are evaluated from propagation of errors,
while the systematic uncertainties are determined as described in
\secRef{sec:systematics}.

\begin{figure}[p]
\includegraphics[width=\textwidth]{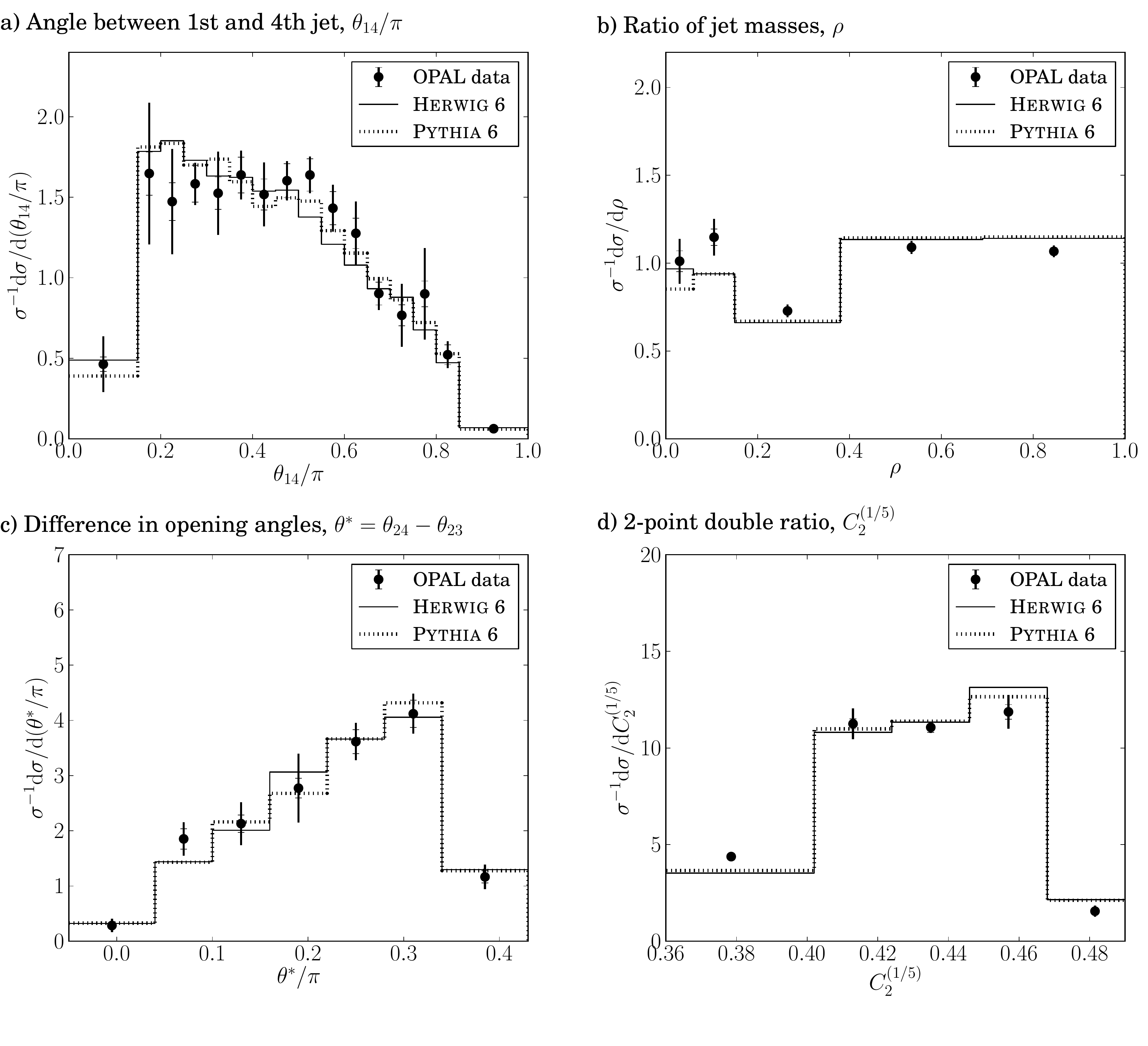}
\caption{The corrected distributions of \tit{a)} the emission angle
  $\theta_{14}$, \tit{b)} the mass ratio $\rho=M_L^2/M_H^2$, \tit{c)}
  the difference in opening angles $\theta^*$, and \tit{d)} the
  2-point double ratio $C_2^{(1/5)}$, in comparison with the
  predictions of the \Hw{6}\ and \Py{6}\ Monte Carlo event generators
  at the hadron level. The error bars limited by the horizontal lines
  indicate the statistical uncertainties, while the total
  uncertainties correspond to the full error bars.  }
\label{fig:data}
\end{figure}

\afterpage{
\begin{landscape}

\begin{table}[p]
\begin{center}
{\small{
\begin{tabular}{cr}
\toprule
$\theta_{14}/\pi$ & $\sigma^{-1}~\mrm{d}\sigma/\mrm{d}(\theta^*/\pi)$ \\
\midrule
$0.00-0.15$ & $0.4631\pm0.0454\pm0.1670$ \\
$0.15-0.20$ & $1.6474\pm0.1345\pm0.4193$ \\
$0.20-0.25$ & $1.4728\pm0.1172\pm0.3055$ \\
$0.25-0.30$ & $1.5833\pm0.1139\pm0.0664$ \\
$0.30-0.35$ & $1.5249\pm0.1004\pm0.2382$ \\
$0.35-0.40$ & $1.6383\pm0.1109\pm0.1032$ \\
$0.40-0.45$ & $1.5172\pm0.0964\pm0.1741$ \\
$0.45-0.50$ & $1.6025\pm0.1053\pm0.0624$ \\
\bottomrule 
\end{tabular} \hspace*{1mm}
\begin{tabular}{cr}
 & \vspace*{1.3mm} \\
\midrule
$0.50-0.55$ & $1.6381\pm0.1000\pm0.0557$ \\
$0.55-0.60$ & $1.4319\pm0.1024\pm0.1045$ \\
$0.60-0.65$ & $1.2758\pm0.0945\pm0.1736$ \\
$0.65-0.70$ & $0.9020\pm0.0708\pm0.0741$ \\
$0.70-0.75$ & $0.7668\pm0.0648\pm0.1853$ \\
$0.75-0.80$ & $0.8999\pm0.0798\pm0.2725$ \\
$0.80-0.85$ & $0.5220\pm0.0616\pm0.0566$ \\
$0.85-1.00$ & $0.0626\pm0.0087\pm0.0223$ \\
\bottomrule 
\end{tabular}
\vspace*{4mm} \\
\begin{tabular}{rrrrrrrrrrrrrrrr}
\toprule
\multicolumn{6}{l}{Correlation Matrix} \\
\midrule
$1.000$  \\
$0.252$ & $1.000$  \\
$-0.037$ & $0.105$ & $1.000$  \\
$-0.048$ & $-0.062$ & $0.103$ & $1.000$  \\
$-0.068$ & $-0.081$ & $-0.108$ & $0.190$ & $1.000$  \\
$-0.055$ & $-0.064$ & $-0.087$ & $-0.082$ & $0.141$ & $1.000$  \\
$-0.068$ & $-0.077$ & $-0.082$ & $-0.118$ & $-0.106$ & $0.164$ & $1.000$  \\
$-0.055$ & $-0.072$ & $-0.084$ & $-0.081$ & $-0.114$ & $-0.105$ & $0.153$ & $1.000$  \\
$-0.073$ & $-0.071$ & $-0.099$ & $-0.093$ & $-0.113$ & $-0.104$ & $-0.123$ & $0.273$ & $1.000$  \\
$-0.068$ & $-0.068$ & $-0.078$ & $-0.078$ & $-0.088$ & $-0.085$ & $-0.111$ & $-0.100$ & $0.292$ & $1.000$  \\
$-0.072$ & $-0.065$ & $-0.083$ & $-0.086$ & $-0.097$ & $-0.089$ & $-0.095$ & $-0.111$ & $-0.090$ & $0.181$ & $1.000$  \\
$-0.066$ & $-0.061$ & $-0.092$ & $-0.096$ & $-0.091$ & $-0.062$ & $-0.096$ & $-0.094$ & $-0.130$ & $-0.097$ & $0.290$ & $1.000$  \\
$-0.050$ & $-0.074$ & $-0.092$ & $-0.088$ & $-0.093$ & $-0.038$ & $-0.112$ & $-0.108$ & $-0.098$ & $-0.109$ & $-0.110$ & $0.273$ & $1.000$  \\
$-0.073$ & $-0.084$ & $-0.091$ & $-0.091$ & $-0.044$ & $-0.116$ & $-0.129$ & $-0.099$ & $-0.107$ & $-0.110$ & $-0.116$ & $-0.108$ & $0.259$ & $1.000$  \\
$-0.075$ & $0.003$ & $-0.038$ & $-0.073$ & $-0.105$ & $-0.078$ & $-0.103$ & $-0.099$ & $-0.087$ & $-0.073$ & $-0.096$ & $-0.090$ & $-0.079$ & $0.179$ & $1.000$  \\
$-0.072$ & $0.006$ & $-0.091$ & $-0.092$ & $-0.088$ & $-0.091$ & $-0.090$ & $-0.045$ & $-0.094$ & $-0.031$ & $-0.095$ & $-0.104$ & $-0.092$ & $0.055$ & $0.190$ & $1.000$ \\
\bottomrule 
\end{tabular}
}}
\end{center}
\caption{The normalized corrected data and the correlation matrix at
  the hadron level for the emission angle $\theta_{14}$. The first
  uncertainty is statistical and the second systematic.}
\label{tab:dataBinNorm1}
\end{table}

\begin{table}[p]
\begin{center}
{\small{
\begin{tabular}{ll}
\begin{tabular}{cr}
\toprule
$\rho$ & $\sigma^{-1}~\mrm{d}\sigma/\mrm{d}\rho$ \\
\midrule
$0.00-0.06$ & $1.0108\pm0.0589\pm0.1137$ \\
$0.06-0.15$ & $1.1474\pm0.0470\pm0.0934$ \\
$0.15-0.38$ & $0.7278\pm0.0190\pm0.0308$ \\
$0.38-0.69$ & $1.0901\pm0.0151\pm0.0344$ \\
$0.69-1.00$ & $1.0670\pm0.0207\pm0.0254$ \\
\bottomrule 
\end{tabular}
&
\begin{tabular}{rrrrr}
\toprule
\multicolumn{5}{l}{Correlation Matrix} \\
\midrule
$1.000$  \\
$0.435$ & $1.000$  \\
$-0.186$ & $-0.007$ & $1.000$  \\
$-0.340$ & $-0.530$ & $0.088$ & $1.000$  \\
$-0.193$ & $-0.356$ & $-0.490$ & $0.387$ & $1.000$ \\
\bottomrule 
\end{tabular}
\vspace*{6mm} \\
\begin{tabular}{cr}
\toprule
$\theta^*/\pi$ & $\sigma^{-1}~\mrm{d}\sigma/\mrm{d}(\theta^*/\pi)$ \\
\midrule
$-0.05-0.04$ & $0.2849\pm0.0497\pm0.1101$ \\
$0.04-0.10$ & $1.8538\pm0.1850\pm0.2395$ \\
$0.10-0.16$ & $2.1298\pm0.1579\pm0.3549$ \\
$0.16-0.22$ & $2.7726\pm0.1791\pm0.5958$ \\
$0.22-0.28$ & $3.6151\pm0.2158\pm0.2605$ \\
$0.28-0.34$ & $4.1203\pm0.2477\pm0.2650$ \\
$0.34-0.43$ & $1.1652\pm0.1110\pm0.1946$ \\
\bottomrule 
\end{tabular}
&
\begin{tabular}{rrrrrrr}
\toprule
\multicolumn{7}{l}{Correlation Matrix} \\
\midrule
$1.000$  \\
$-0.013$ & $1.000$  \\
$0.358$ & $0.367$ & $1.000$  \\
$-0.188$ & $0.010$ & $0.232$ & $1.000$  \\
$-0.250$ & $-0.327$ & $-0.092$ & $0.278$ & $1.000$  \\
$-0.219$ & $-0.457$ & $-0.408$ & $-0.295$ & $0.245$ & $1.000$  \\
$-0.191$ & $-0.335$ & $-0.376$ & $-0.382$ & $-0.168$ & $0.304$ & $1.000$ \\
\bottomrule 
\end{tabular}
\vspace*{6mm} \\
\begin{tabular}{cr}
\toprule
$C_2^{(1/5)}$ & $\sigma^{-1}~\mrm{d}\sigma/\mrm{d}C_2^{(1/5)}$ \\
\midrule
$0.355-0.402$ & $4.3798\pm0.1680\pm0.1447$ \\
$0.402-0.424$ & $11.2510\pm0.2603\pm0.7473$ \\
$0.424-0.446$ & $11.0688\pm0.2557\pm0.1536$ \\
$0.446-0.468$ & $11.8690\pm0.3771\pm0.7888$ \\
$0.468-0.495$ & $1.5553\pm0.1137\pm0.2591$ \\
\bottomrule 
\end{tabular}
&
\begin{tabular}{rrrrr}
\toprule
\multicolumn{5}{l}{Correlation Matrix} \\
\midrule
$1.000$  \\
$0.314$ & $1.000$  \\
$-0.385$ & $0.218$ & $1.000$  \\
$-0.458$ & $-0.493$ & $0.187$ & $1.000$  \\
$-0.260$ & $-0.387$ & $-0.225$ & $0.299$ & $1.000$ \\
\bottomrule 
\end{tabular}
\end{tabular}
}}
\end{center}
\caption{The normalized corrected data and the correlation matrix at
  the hadron level for the mass ratio $\rho=M_L^2/M_H^2$, the
  difference in emission angles $\theta^*$, and the 2-point double
  ratio $C_2^{(1/5)}$.  The first uncertainty is statistical and the
  second systematic.}
\label{tab:dataBinNorm2}
\end{table}

\begin{table}[p]
{\small{
\begin{center}
\begin{tabular}{cr}
\toprule
\# & Central/Towards \\
\midrule
$1$ & $1.0159\pm0.0535\pm0.1059$ \\
$2$ & $2.0383\pm0.1447\pm0.7756$ \\
$3$ & $0.6687\pm0.0304\pm0.0271$ \\
$4$ & $0.5319\pm0.0324\pm0.0432$ \\
$5$ & $1.0671\pm0.0825\pm0.3804$ \\
$6$ & $0.3501\pm0.0192\pm0.0139$ \\
$7$ & $1.4942\pm0.0740\pm0.1579$ \\
$8$ & $2.9979\pm0.2060\pm1.1161$ \\
$9$ & $0.9836\pm0.0411\pm0.0339$ \\
\bottomrule
\end{tabular} \hspace*{5mm}
\begin{tabular}{cr}
\toprule
\# & Central/Away \\
\midrule
$1$ & $1.3591\pm0.0675\pm0.0939$ \\
$2$ & $2.6045\pm0.1591\pm0.4134$ \\
$3$ & $8.7205\pm0.8699\pm1.2979$ \\
$4$ & $0.7116\pm0.0415\pm0.0618$ \\
$5$ & $1.3636\pm0.0932\pm0.2226$ \\
$6$ & $4.5656\pm0.4765\pm0.7924$ \\
$7$ & $1.9989\pm0.0926\pm0.1683$ \\
$8$ & $3.8307\pm0.2238\pm0.6735$ \\
$9$ & $12.8260\pm1.2589\pm1.9634$ \\
\bottomrule
\end{tabular} \hspace*{5mm}
\begin{tabular}{cr}
\toprule
\# & Towards/Away \\
\midrule
$1$ & $1.3378\pm0.0745\pm0.2070$ \\
$2$ & $2.5637\pm0.1695\pm0.5789$ \\
$3$ & $8.5840\pm0.8834\pm2.0061$ \\
$4$ & $0.6668\pm0.0489\pm0.2315$ \\
$5$ & $1.2778\pm0.1041\pm0.5149$ \\
$6$ & $4.2784\pm0.4851\pm1.7585$ \\
$7$ & $2.0323\pm0.0995\pm0.2145$ \\
$8$ & $3.8947\pm0.2358\pm0.7391$ \\
$9$ & $13.0404\pm1.2966\pm2.3652$ \\
\bottomrule
\end{tabular} \vspace*{5mm} \\ 
\begin{tabular}{cr}
\toprule
$\theta^*_0$ & $N_\text{left}/N_\text{right}$ \\
\midrule
$0.10$ & $0.1586\pm0.0147\pm0.0268$ \\
$0.16$ & $0.3599\pm0.0240\pm0.0179$ \\
$0.22$ & $0.7575\pm0.0441\pm0.1196$ \\
$0.28$ & $1.8402\pm0.1137\pm0.2619$ \\
\bottomrule
\end{tabular}  \hspace*{5mm} 
\begin{tabular}{cr}
\toprule
$\rho_0$ & $N_\text{left}/N_\text{right}$ \\
\midrule
$0.15$ & $0.1961\pm0.0069\pm0.0212$ \\
$0.38$ & $0.4955\pm0.0121\pm0.0312$ \\
\bottomrule
 & \\
 & \\
\end{tabular} \hspace*{5mm}
\begin{tabular}{cr}
\toprule
$C_{2,0}^{(1/5)}$ & $N_\text{left}/N_\text{right}$ \\
\midrule
$0.42$ & $0.8294\pm0.0239\pm0.0503$ \\
$0.45$ & $2.2991\pm0.0767\pm0.1346$ \\
\bottomrule
 & \\
 & \\
\end{tabular} \hspace*{5mm}
\end{center}
\caption{The corrected data for the derived distributions. The upper
  three tables list the results for $\theta_{14}$ asymmetry ratios
  defined in \EqRef{eq:A14ratios}, with the definitions of the
  towards, central, and away regions given in
  \TabRef{tab:DefRegionsA14}. The bottom three tables list the results
  for the asymmetries defined for the other observables.  The first
  uncertainty is statistical and the second systematic}
\label{tab:dataAS}

\vspace*{5mm}

\begin{center}
\begin{tabular}{lrrrrrr}
\toprule & \hpp $\tilde q^2$ & \hpp $p_{\perp\mrm{dip}}^2$ & \hpp $q_\mrm{dip}^2$ & \Py{8}\ $p_{\perp\mrm{evol}}^2$
 & \vc $p_{\perp\mrm{ant}}^2$ & \vc $m_\mrm{ant}^2$ \\
\midrule
$\theta_{14}/\pi$   & $4.8~~(9.968\times10^{-1})$  & $12.8~~(6.850\times10^{-1})$ & $16.4~~(4.230\times10^{-1})$ & $8.3~~(9.401\times10^{-1})$ & $8.1~~(9.718\times10^{-1})$ & $7.1~~(9.444\times10^{-1})$ \\ 
$\theta^*/\pi$      & $6.1~~(5.243\times10^{-1})$  & $2.8~~(9.032\times10^{-1})$  & $8.8~~(2.683\times10^{-1})$  & $2.5~~(9.239\times10^{-1})$ & $3.3~~(6.496\times10^{-1})$ & $5.1~~(8.552\times10^{-1})$ \\ 
$C_2^{(1/5)}$       & $16.2~~(6.196\times10^{-3})$ & $6.0~~(3.083\times10^{-1})$  & $41.2~~(8.498\times10^{-8})$ & $4.3~~(5.090\times10^{-1})$ & $5.3~~(8.013\times10^{-1})$ & $2.3~~(3.753\times10^{-1})$ \\ 
$\rho$              & $6.6~~(2.496\times10^{-1})$  & $10.1~~(7.15\times10^{-2})$ & $42.8~~(4.098\times10^{-8})$ & $2.6~~(7.633\times10^{-1})$ & $6.8~~(5.682\times10^{-1})$ & $3.9~~(2.354\times10^{-1})$ \\ 
$\langle p \rangle$ & $~~~~~(4.442\times10^{-1})$  &  $~~~~~(4.92\times10^{-1})$ & $~~~~~(1.728\times10^{-1})$  & $~~~~~(7.841\times10^{-1})$ & $~~~~~(7.477\times10^{-1})$ & $~~~~~(6.026\times10^{-1})$ \\
\bottomrule
\end{tabular}
\end{center}
\caption{The $\chi^2$ values (with p-values in parentheses) for the
  four observables considered in this analysis. The last line gives
  the average p-value, $\langle p \rangle$, for the four observables.}
\label{tab:chi2values}
}}
\end{table}

\end{landscape}
}

\subsection{Systematic uncertainties \label{sec:systematics}}

Systematic uncertainties are evaluated by repeating the analysis with
different selection requirements and with variations in the correction
procedure. Specifically, we consider the following:
\begin{itemize}
\item The requirement on the thrust angle direction is changed to
  $|\cos\theta_\mrm{T}|<0.7$ from the default
  $|\cos\theta_\mrm{T}|<0.9$.
\item The minimum number of charged tracks is increased to seven from
  the default of five.
\item Variation of the reconstruction procedure: All tracks and
  clusters are taken into account. In this case the detector
  correction takes care of the double counting.
\item \Hw{6}\ is used in place of \Py{6} to determine the response matrix.
\end{itemize}
The systematic uncertainty is determined for each variation from the
bin-by-bin difference in the corrected distributions with respect to
the standard result.  The total systematic uncertainty is given by the
quadrature sum of the individual terms. The total uncertainty of the
data is defined by summing the statistical and systematic
contributions in quadrature.

As additional systematic checks on the unfolding procedure, we
consider the following variations:
\begin{itemize}
\item We use the unfolding method with three and five instead of four
  iterations.
\item Instead of the iterative method, we use the singular value
  decomposition, as proposed by H{\"o}cker and
  Kartvelishvili~\cite{Hocker:1995kb} and implemented in \roou.
\end{itemize}
We find the systematic variations that arise from these two checks to
be smaller or comparable to the variation observed when using
\Hw{6}\ in place of \Py{6}. Since adding all these effects together
would likely double count the uncertainty associated with the
unfolding procedure we do not add the observed differences to the
systematic uncertainty.

\section{Comparison with Monte Carlo models \label{sec:comp}}

In this section, we present a comparison between the coherence schemes
described in \secRef{sec:models} and the data. For this purpose,
samples of $5\times 10^6$ events are generated for each MC model,
using the tuned parameter sets mentioned in \secRef{sec:models}. We
present the predictions for the different schemes in terms of the
observables defined in \secRef{sec:intro}.  As a measure of the level
of agreement with data, we calculate the significance, defined as
\begin{align}
\sigma_i = \frac{\text{MC}_i-\mathcal D_i}{\sigma_{\mathcal D_i}}~,
\end{align}
where $\text{MC}_i$ and $\mathcal D_i$ represent the predicted and
observed values in bin $i$ of a distribution, with $\sigma_{\mathcal
  D_i}$ the corresponding uncertainty in $\mathcal D_i$.  In the
following, we present a distribution of the significance in a plot
below the distribution of the variables.  In addition we calculate the
$\chi^2$ values for the distribution as
\begin{align}
\chi^2 = (\text{MC}-\mathcal D)^{\text T}V^{-1}(\text{MC}-\mathcal D)
=\sum_{i,j=1}^{N_\mrm{bins}}(\text{MC}-\mathcal D)_i(V^{-1})_{ij}(\text{MC}-\mathcal D)_j
~,
\end{align}
where $V$ is the full covariance matrix, representing statistical
terms, with systematic uncertainties added to the diagonal elements.
We present the $\chi^2$ results and corresponding p-values in
\TabRef{tab:chi2values}.  The latter are calculated with the
\roo~\cite{Brun:1997pa} program and give the probability that the
deviations of the MC predictions from the data are consistent with the
evaluated uncertainties.

\subsection{Angle between first and fourth jet: $\theta_{14}$}

\begin{figure}[t]
\centering
\includegraphics[width=\textwidth]{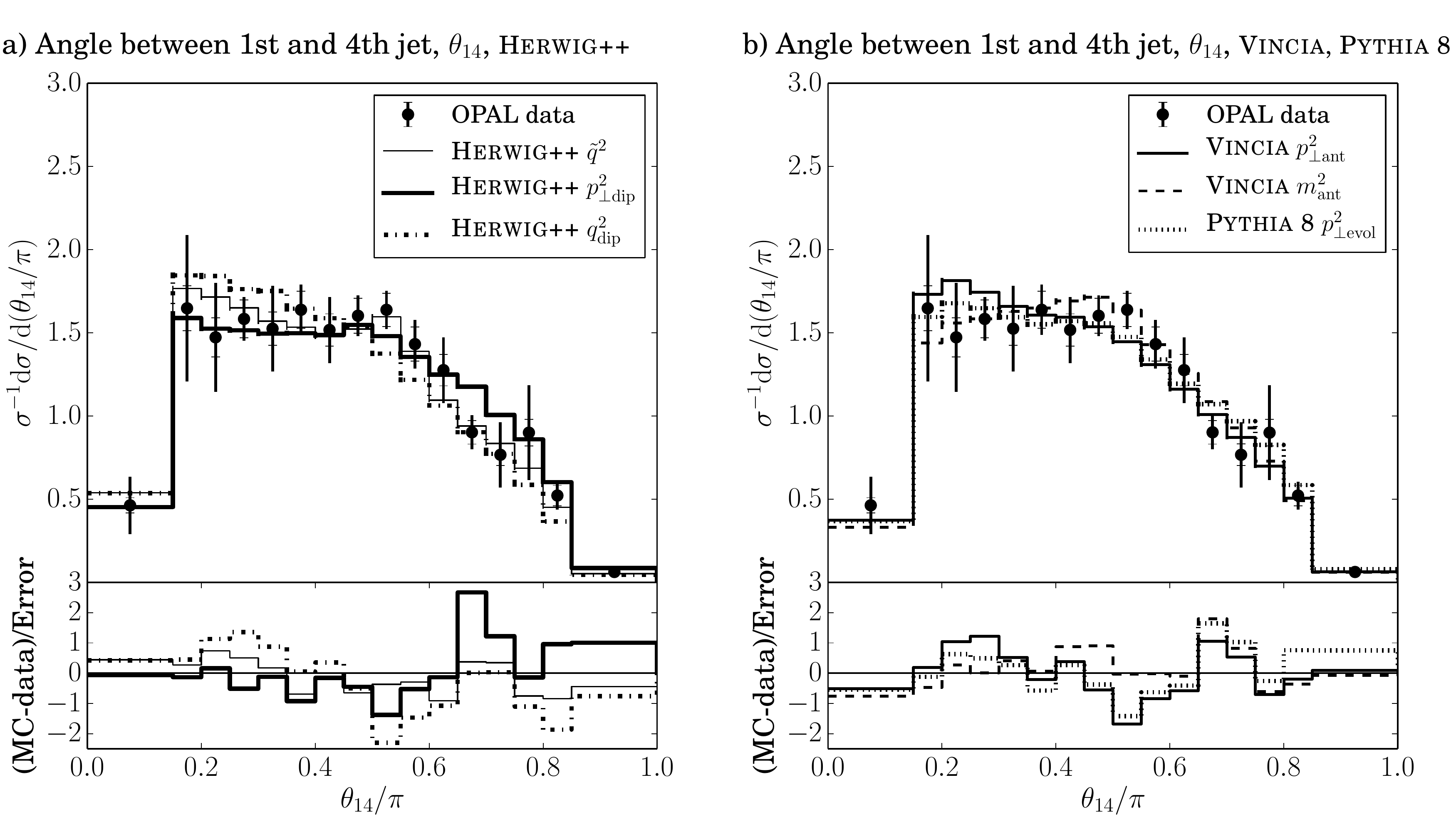}
\caption{The corrected distribution of the emission angle
  $\theta_{14}$ of the soft fourth jet in comparison with the
  predictions of \tit{a)} \hpp and \tit{b)} \Py{8}\ and \vc.  The thin
  solid lines correspond to \hpp with angular-ordering ($\tilde q^2$), the thick
  solid lines to the dipole shower of \hpp with ordering in
  $p_{\perp\mrm{dip}}^2$, and the dash-dotted lines to ordering in
  $q_\mrm{dip}^2$.  \vc with ordering in $p_{\perp\mrm{ant}}^2$ is
  shown with medium solid lines, ordering in $m_\mrm{ant}^2$ with
  dashed lines, and \Py{8}\ is shown with dotted lines.  The error
  bars limited by the horizontal lines indicate the statistical
  uncertainties, while the total uncertainties correspond to the full
  error bars.  The ratio plots show the deviation of the predictions
  from the data in units of the total uncertainty.}
\label{fig:angle14data}
\end{figure}

In \FigsRef{fig:angle14data} \tit{a)} and \tit{b)} we show the
normalized distribution of the emission angle of the soft fourth jet
from the nearly collinear three-jet system, $\theta_{14}$.  All models
are found to provide adequate descriptions of the data, except that
the \hpp $p_{\perp\mrm{dip}}^2$ model lies about three standard deviations
above the measurements for a narrow region around $\theta_{14}\approx
0.7\pi$. Even for this model, the p-value is $0.42$
(\TabRef{tab:chi2values}), implying overall compatibility with the
experimental results.

\begin{figure}[p]
\centering
\includegraphics[scale=0.42]{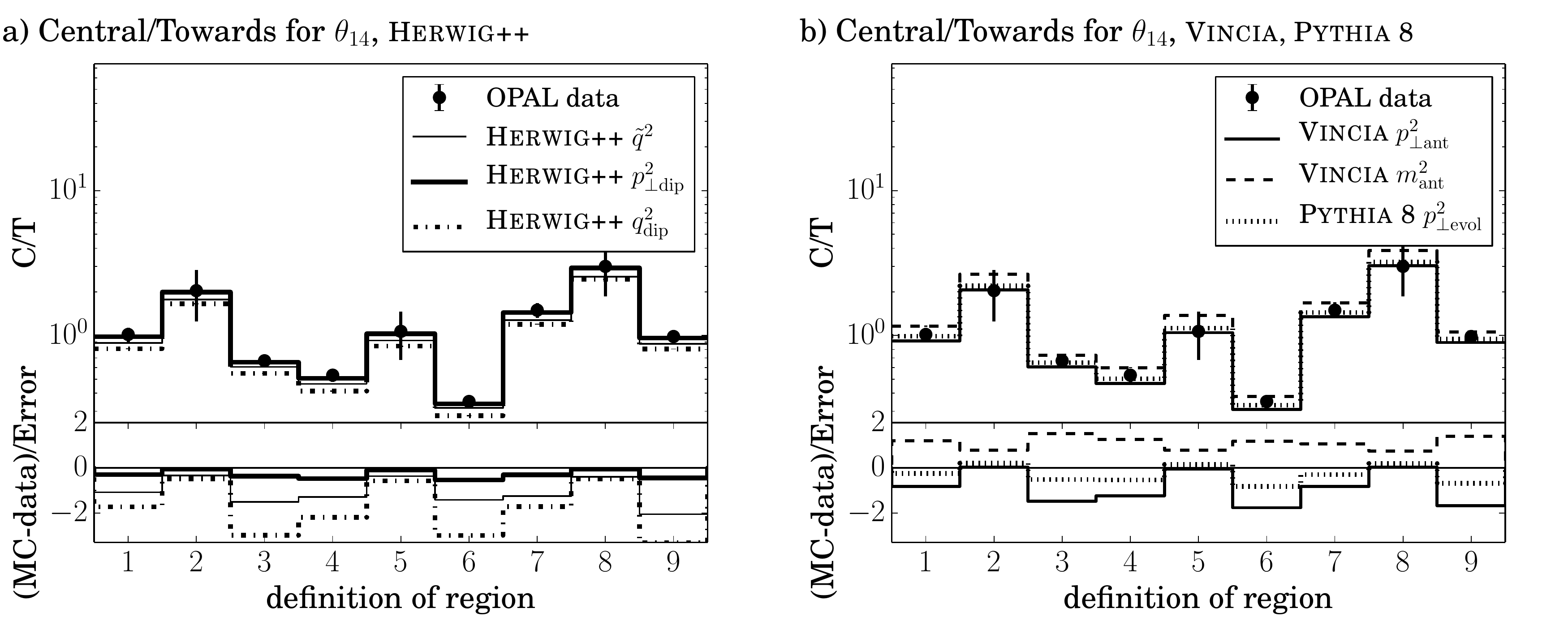} \\
\includegraphics[scale=0.42]{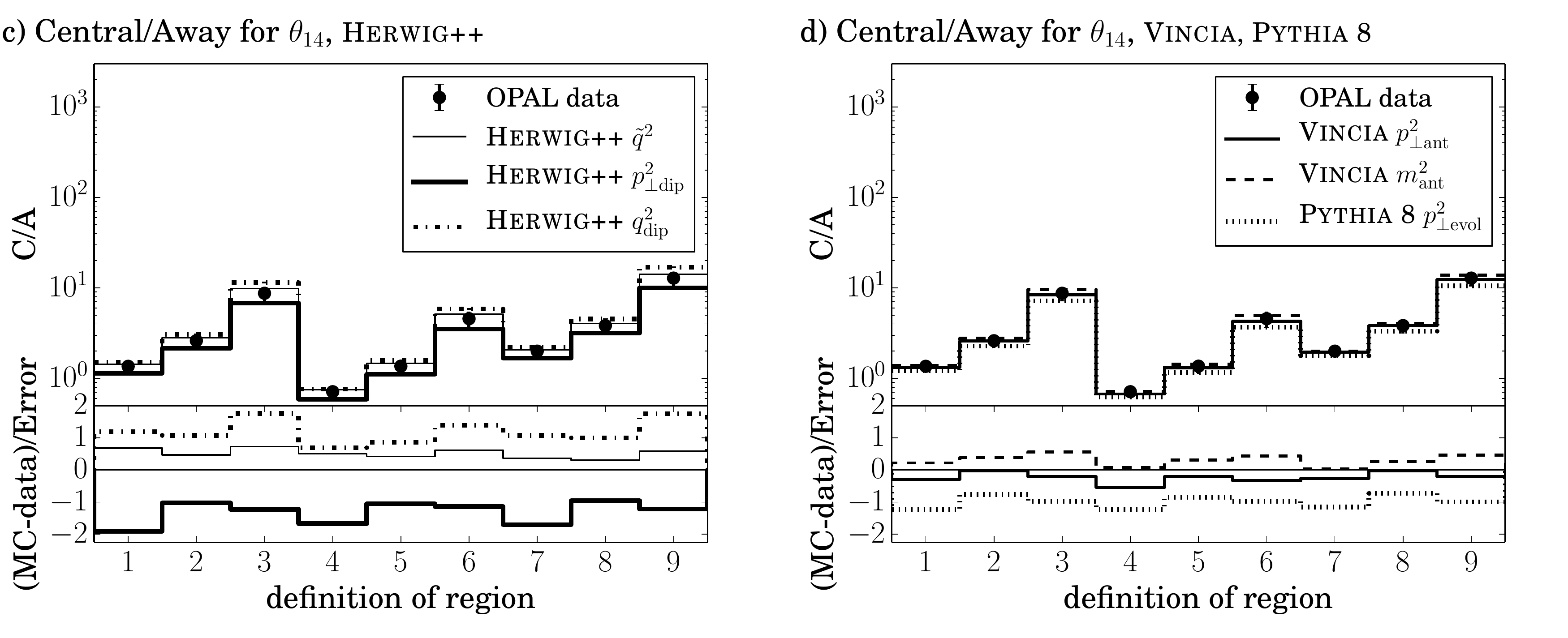} \\
\includegraphics[scale=0.42]{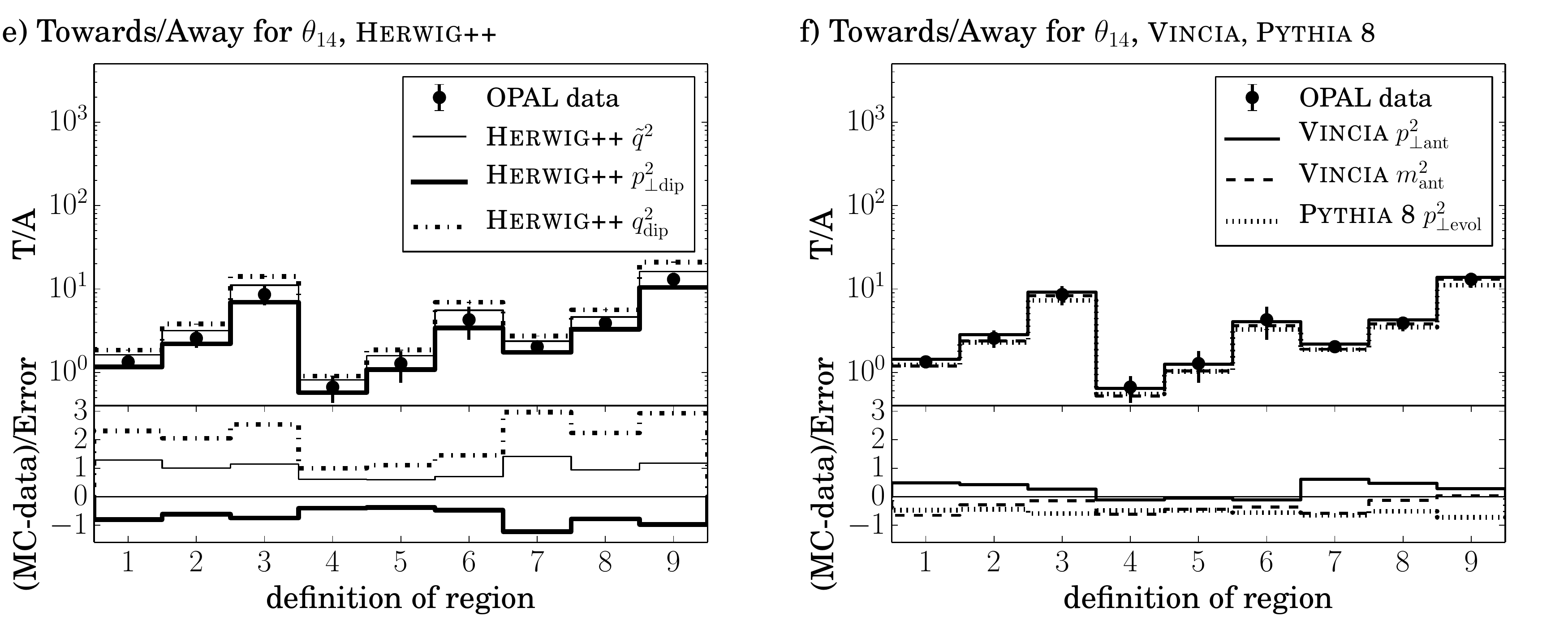}
\caption{The corrected data for the derived distributions in
  comparison with the predictions of \tit{a)} \hpp and \tit{b)}
  \Py{8}\ and \vc.  The thin solid lines correspond to \hpp with
  angular-ordering ($\tilde q^2$), the thick solid lines to the dipole shower of \hpp
  with ordering in $p_{\perp\mrm{dip}}^2$, and the dash-dotted lines
  to ordering in $q_\mrm{dip}^2$.  \vc with ordering in
  $p_{\perp\mrm{ant}}^2$ is shown with medium solid lines, ordering in
  $m_\mrm{ant}^2$ with dashed lines, and \Py{8}\ is shown with dotted
  lines.  The error bars limited by the horizontal lines indicate the
  statistical uncertainties, while the total uncertainties correspond
  to the full error bars.  The ratio plots show the deviation of the
  predictions from the data in units of the total uncertainty.}
\label{fig:angle14dataAS}
\end{figure}

We show the ratio C/T of the central-to-towards regions, which gives
the relative amount of wide-angle to collinear emissions, in
\FigsRef{fig:angle14dataAS}\tit{a)} and \tit{b)}. For the
$p_{\perp\mrm{dip}}^2$-ordered dipole shower of \hpp and the parton shower
of \Py8\ we find nearly perfect agreement with the data for all nine
C/T regions (\TabRef{tab:DefRegionsA14}). The \hpp $\tilde q^2$ model
and the \vc $p^2_{\perp\mrm{ant}}$ model lie below the data by up to
two standard deviations in some regions, while the \vc
$m^2_\mrm{ant}$ model lies about two standard deviations above
the data in all regions. The two Vincia models exhibit the expected
behavior: When the antenna mass is used as the evolution variable,
soft wide-angle emissions are preferred over collinear ones, which
leads to higher values for the relative level of wide-angle to
collinear emissions. This demonstrates the sensitivity of the
$\theta_{14}$ variable to the choice of evolution scheme.  The largest
deviation from the data in \FigsRef{fig:angle14dataAS}\tit{a)} and
\tit{b)} is observed for the $q_{\mrm{dip}}^2$-ordered dipole shower
of \hpp, for which the predictions lie up to around three standard
deviations below the data in some regions.  Thus, this model predicts
too many collinear emissions compared to wide-angle emissions.

In \FigsRef{fig:angle14dataAS} \tit{c)} and \tit{d)} we show a
comparison of the MC predictions to the data for the ratio C/A of the
central-to-away regions.  This ratio measures the relative amount of
wide-angle emissions to emissions in a backwards direction, away from
the leading jet and near to the collinear (23) jet pair. For the \hpp
$\tilde q^2$ model and for \vc, we find a good agreement with the data
and observe small differences for the different evolution variables 
of \vc. \Py8\ and the $p_{\perp\mrm{dip}}^2$-ordered dipole shower of 
\hpp lie below the data, by around one and two standard deviations, 
respectively, and thus predict too few wide-angle emissions compared 
to the backwards emissions. The \hpp $q_{\mrm{dip}}^2$ model lies 
around one standard deviation above the data and thus predicts relatively 
too many wide-angle emissions. The observations the ratios C/T and C/A 
are confirmed by the measurements of the ratio T/A of the towards-to-away 
regions, presented in \FigsRef{fig:angle14dataAS} \tit{e)} and \tit{f)}.

\subsection{Difference in opening angles: $\theta^*$}

In \FigsRef{fig:anglesdata} \tit{a)} and \tit{b)} we show the
normalized distribution of the difference in opening angles between
the third and the fourth jet with respect to the second jet,
$\theta^*=\theta_{24}-\theta_{23}$.  All models are seen to provide an
adequate description of the data, with the exception of the region
around $\theta^*\approx 0.07\pi$ (second bin of
\FigsRef{fig:anglesdata} \tit{a)} and \tit{b)}), where the models
predict somewhat fewer events than are observed. The largest
discrepancy in this region arises from the \hpp $q_{\mrm{dip}}^2$
model.

\begin{figure}[p]
\centering
\includegraphics[width=\textwidth]{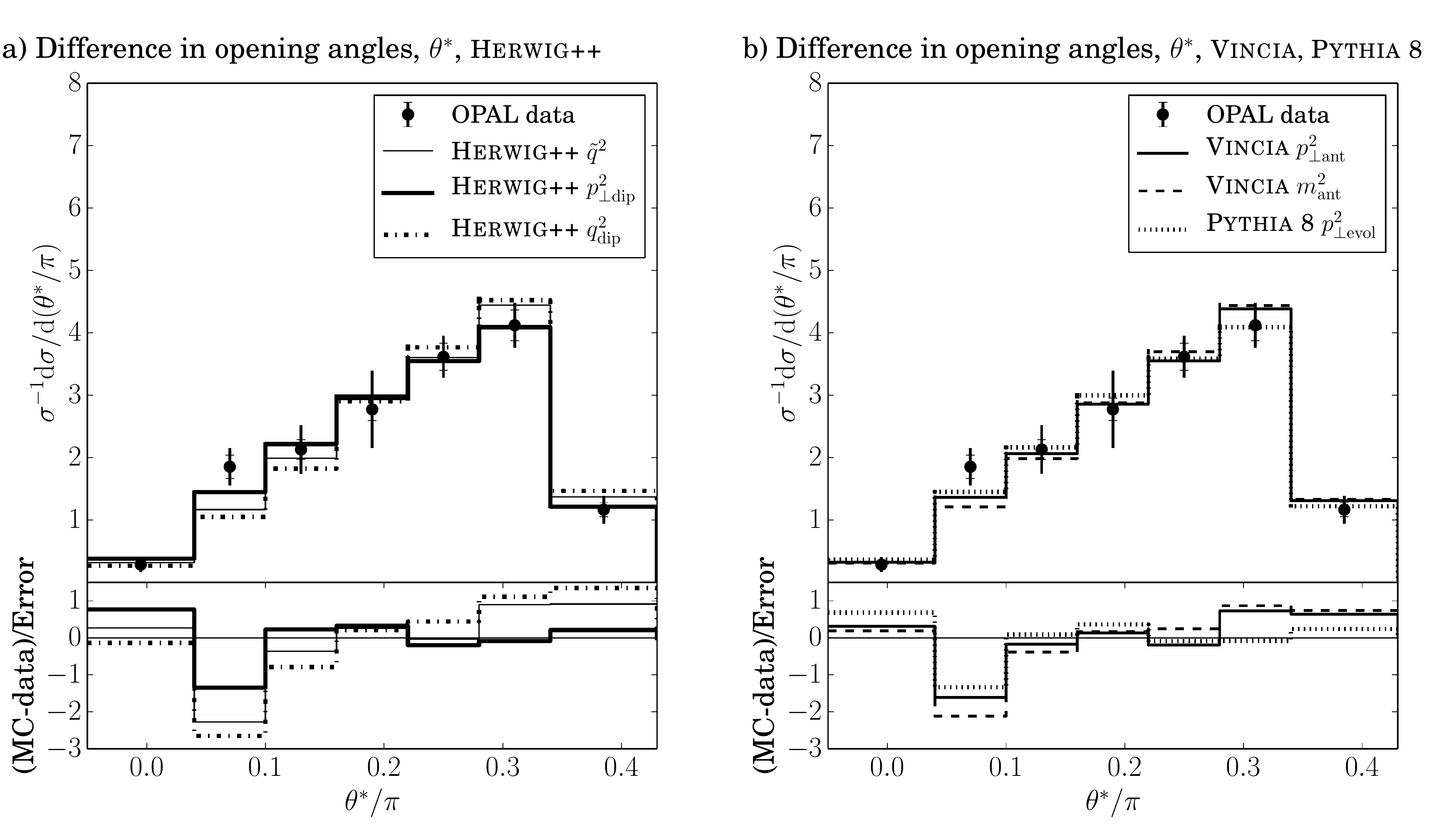} \\
\includegraphics[width=\textwidth]{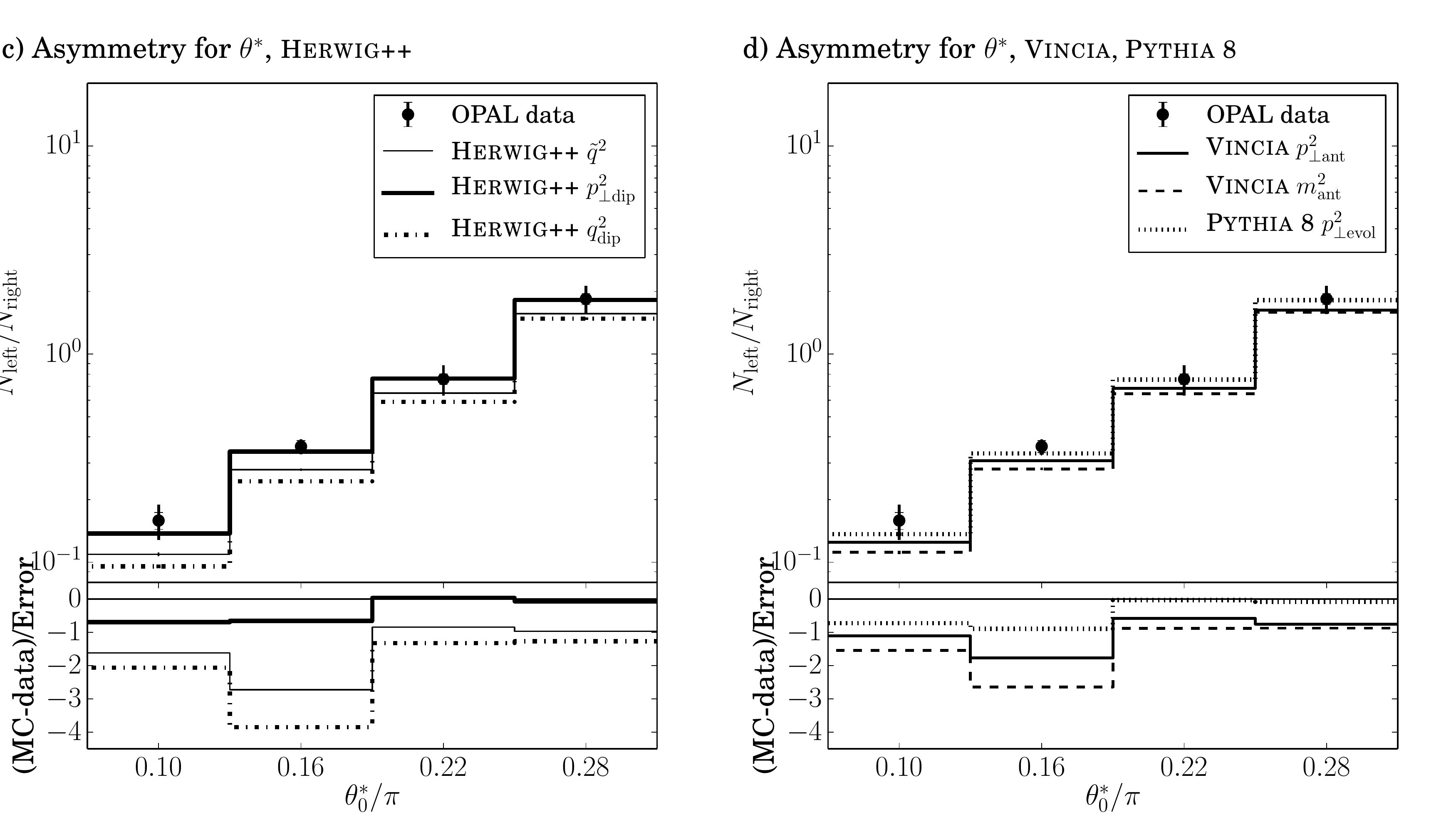} 
\caption{The distribution of the difference in opening angles
  $\theta^*$ for \tit{a)} \hpp and \tit{b)} \Py{8}\ and \vc. The
  asymmetry with respect to the dividing point $\theta^*_0$ is shown
  for \tit{c)} \hpp and \tit{d)} \Py{8}\ and \vc. The thin solid lines
  correspond to \hpp with angular-ordering ($\tilde q^2$), the thick solid lines to
  the dipole shower of \hpp with ordering in $p_{\perp\mrm{dip}}^2$,
  and the dash-dotted lines to ordering in $q_\mrm{dip}^2$.  \vc with
  ordering in $p_{\perp\mrm{ant}}^2$ is shown with medium solid lines,
  ordering in $m_\mrm{ant}^2$ with dashed lines and \Py{8}\ is shown
  with dotted lines.  The error bars limited by the horizontal lines
  indicate the statistical uncertainties, while the total
  uncertainties correspond to the full error bars.  The ratio plots
  show the deviation of the predictions from the data in units of the
  total uncertainty.}
\label{fig:anglesdata}
\end{figure}

We show the asymmetry as a function of the dividing point $\theta^*_0$
in the \FigsRef{fig:anglesdata} \tit{c)} and \tit{d)}. The largest
discriminating power is found for $\theta_0*=0.16\pi$, where the
$q_\mrm{dip}^2$-ordered dipole shower of \hpp generates a deviation of
almost four standard deviations with respect to the data. The number
of events with large differences in the opening angles of the third
and fourth jets is overestimated by this non-coherent shower
model. The $p^2_{\perp\mrm{dip}}$-ordered \hpp shower, based on the
same shower kernels, but respecting coherence due to the choice of
evolution variable, gives a better description of the asymmetry. This
emphasizes the need for coherence in order to describe the data
properly.

\subsection{2-Point double ratio: $C_2^{(1/5)}$}

For the normalized distribution of the 2-point double ratio,
$C_2^{(1/5)}$, shown in \FigsRef{fig:C2data} \tit{a)} and \tit{b)}, we
find rather large deviations between the data and the MC prediction
for most of the shower models. We again find that the
$q_\mrm{dip}^2$-ordered \hpp shower exhibits the largest
discrepancies. Only two models, \Py8\ and the \vc $m_\mrm{ant}^2$
model, yield p-values larger than $50\%$ (\TabRef{tab:chi2values}).

\begin{figure}[p]
\centering
\includegraphics[width=\textwidth]{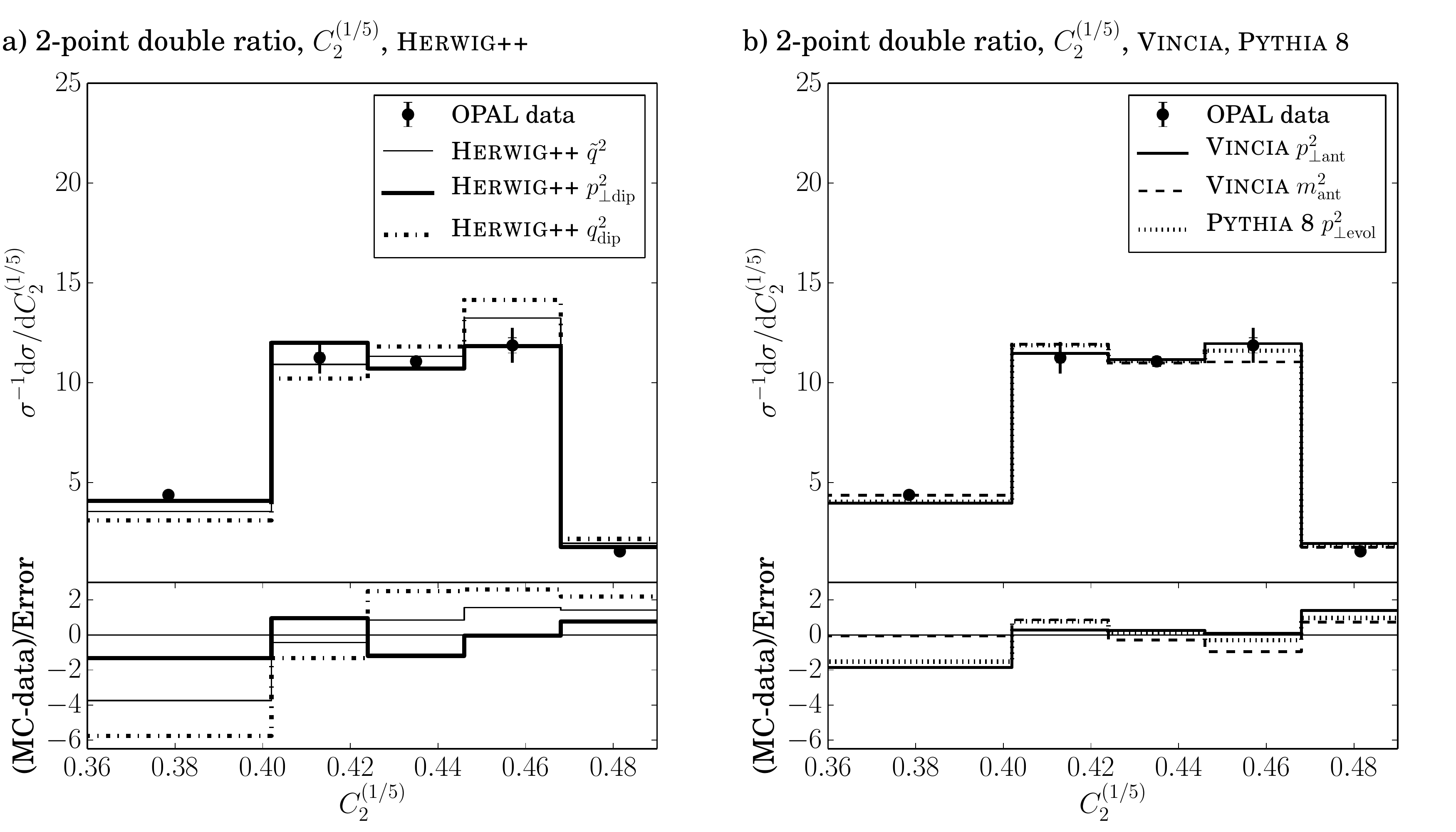} \\
\includegraphics[width=\textwidth]{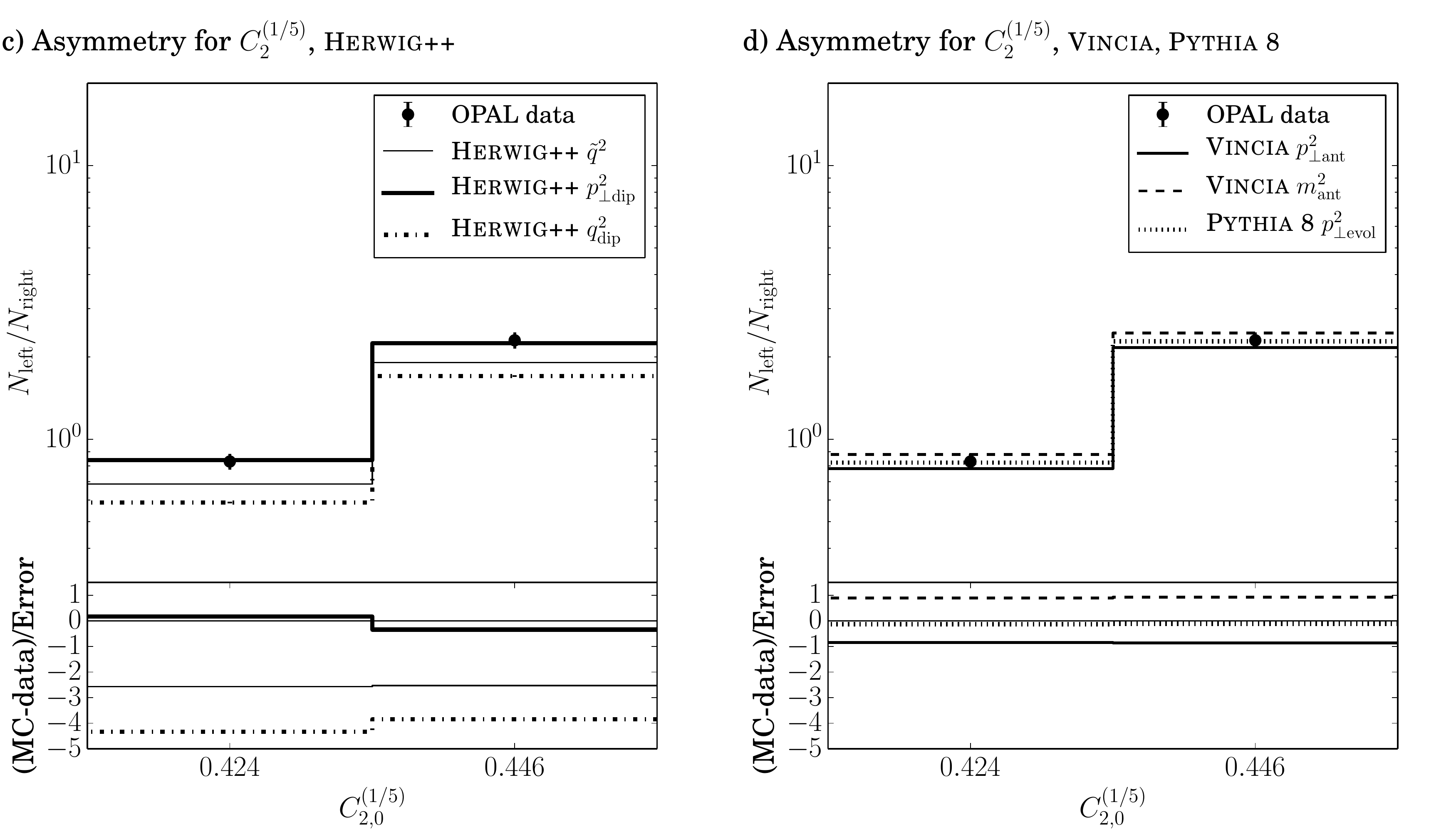}
\caption{The distribution of the difference in opening angles
  $C_2^{(1/5)}$ for \tit{a)} \hpp and \tit{b)} \Py{8}\ and \vc. The
  asymmetry with respect to the dividing point $C_{2,0}^{(1/5)}$ is
  shown for \tit{c)} \hpp and \tit{d)} \Py{8}\ and \vc. The thin solid
  lines correspond to \hpp with angular-ordering ($\tilde q^2$), the thick solid
  lines to the dipole shower of \hpp with ordering in
  $p_{\perp\mrm{dip}}^2$, and the dash-dotted lines to ordering in
  $q_\mrm{dip}^2$.  \vc with ordering in $p_{\perp\mrm{ant}}^2$ is
  shown with medium solid lines, ordering in $m_\mrm{ant}^2$ with
  dashed lines and \Py{8}\ is shown with dotted lines.  The error bars
  limited by the horizontal lines indicate the statistical
  uncertainties, while the total uncertainties correspond to the full
  error bars.  The ratio plots show the deviation of the predictions
  from the data in units of the total uncertainty.}
\label{fig:C2data}
\end{figure}

In \FigsRef{fig:C2data} \tit{c)} and \tit{d)}, we show the asymmetry
in the $C_2^{(1/5)}$ variable as a function of the dividing point
$C_{2,0}^{(1/5)}$. Since $C_2^{(1/5)}$ is proportional to the energy
$E_4$ of the fourth jet, the asymmetry in $C_2^{(1/5)}$ measures the
relative number of events of soft versus hard fourth-jet emissions.
We observe large deviations from the data, at the level of four
standard deviations, for the \hpp $q_\mrm{dip}^2$ model, which
underpredicts the relative fraction of events with a very soft fourth
jet. A similar discrepancy, at the level of around $2.5$ standard
deviations, is observed for the \hpp $\tilde q^2$ model. The two
versions of \vc exhibit deviations of about one standard deviation in
the opposite sense, i.e., \vc $m_\mrm{ant}^2$ somewhat overpredicts the 
level of hard fourth-jet emissions, whereas \vc $p_{\perp\mrm{ant}}^2$
predicts too few hard fourth-jet emissions. In contrast, the 
\hpp $p_{\perp\mrm{dip}}^2$ and \Py8\ models are in nearly perfect 
agreement with the data.

\subsection{Mass ratio: $\rho=M_L^2/M_H^2$}
The normalized distributions of the $\rho=M_L^2/M_H^2$ variable are
shown in \FigsRef{fig:massratiodata} \tit{a)} and \tit{b)}. For the
\Py8\ and the two \vc models, we find reasonable overall agreement
with the data, with differences on the level of two standard
deviations or less. The \hpp\ models demonstrate larger differences,
with discrepancies reaching the level of four standard deviations for
the \hpp $q_\mrm{dip}^2$ model.

\begin{figure}[p]
\centering
\includegraphics[width=\textwidth]{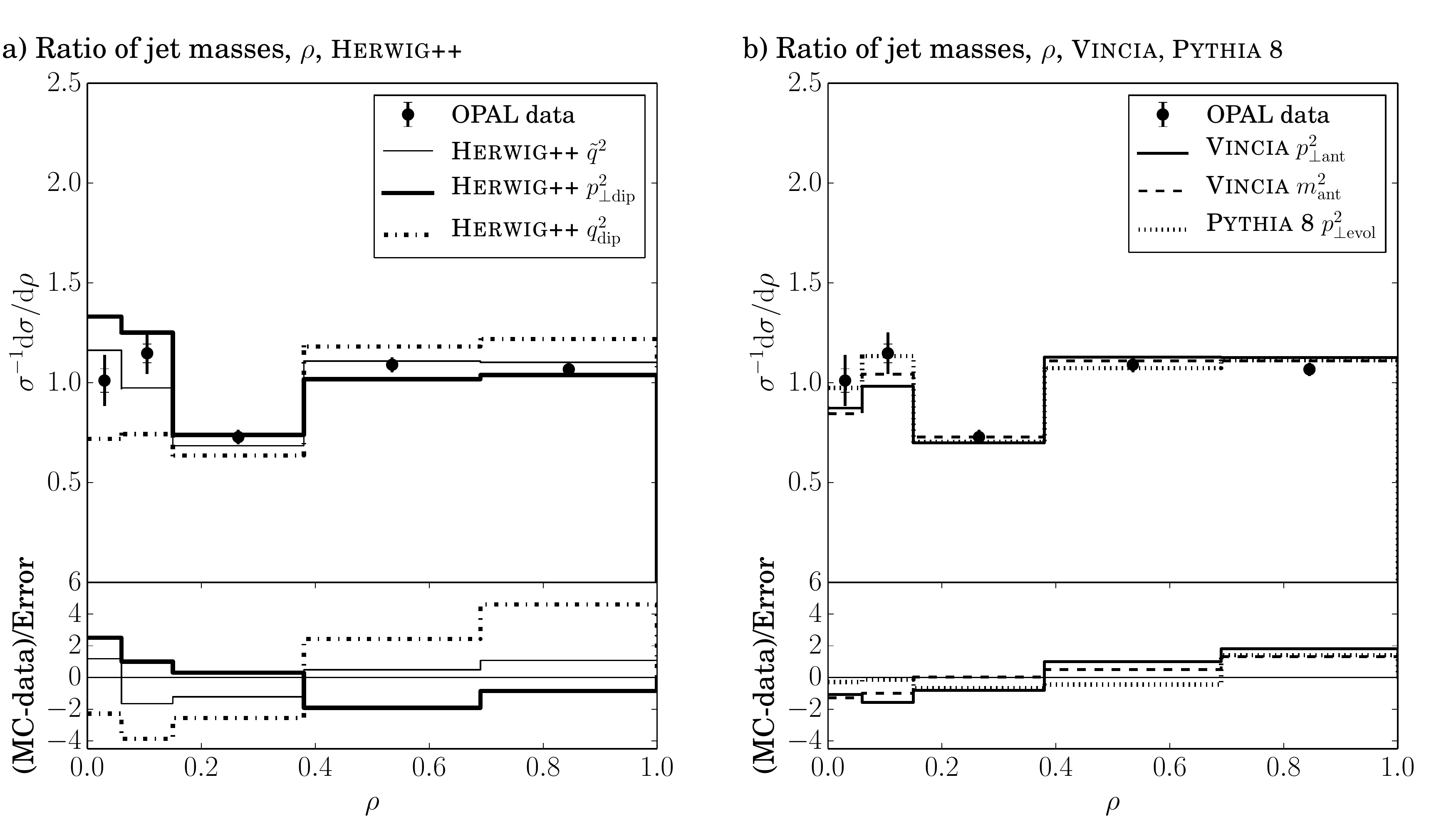} \\
\includegraphics[width=\textwidth]{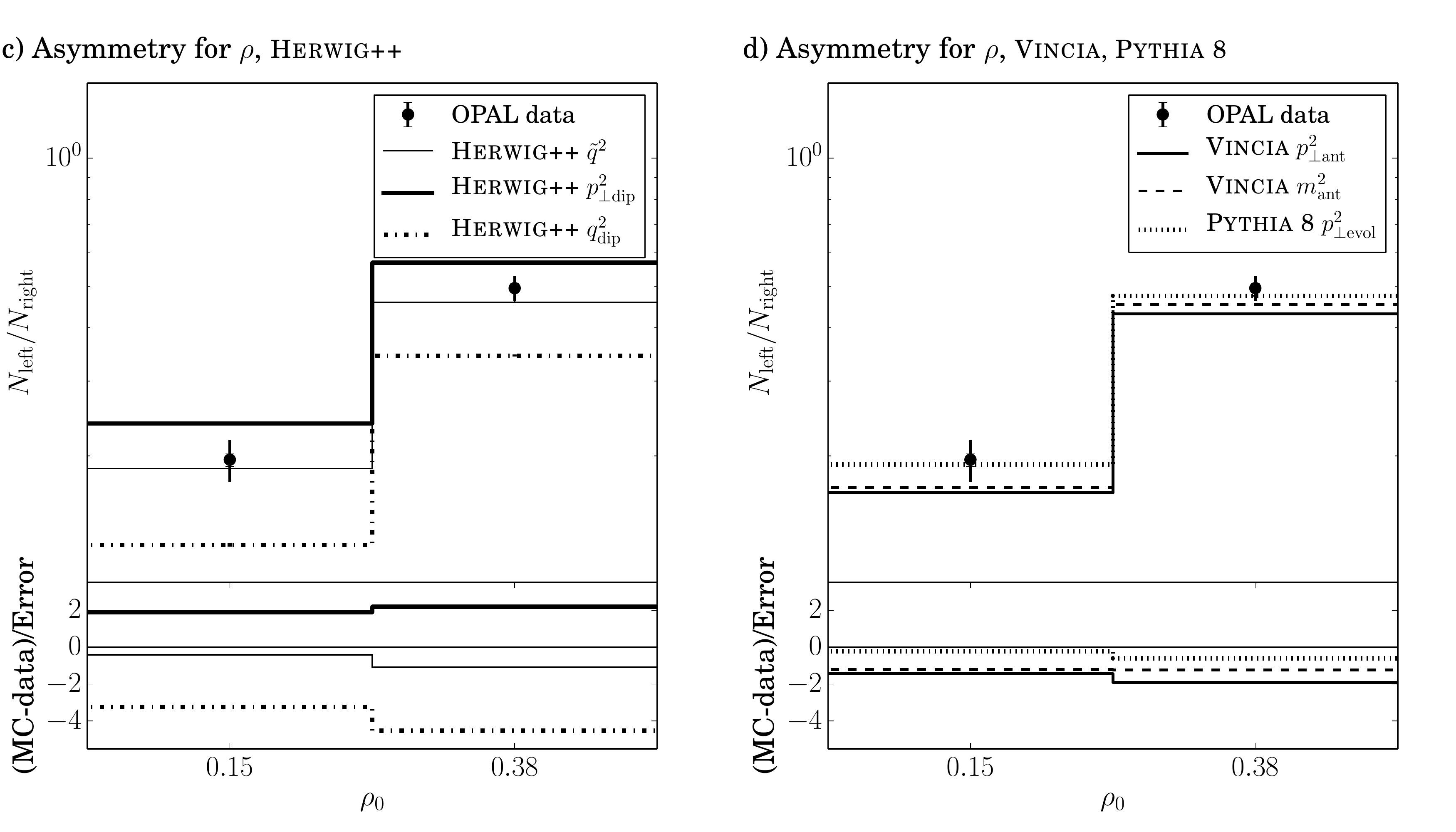}
\caption{The distribution of the difference in opening angles
  $\rho=M_L^2/M_H^2$ for \tit{a)} \hpp and \tit{b)} \Py{8}\ and
  \vc. The asymmetry with respect to the dividing point $\rho_0$ is
  shown for \tit{c)} \hpp and \tit{d)} \Py{8}\ and \vc. The thin solid
  lines correspond to \hpp with angular-ordering ($\tilde q^2$), the thick solid
  lines to the dipole shower of \hpp with ordering in
  $p_{\perp\mrm{dip}}^2$, and the dash-dotted lines to ordering in
  $q_\mrm{dip}^2$.  \vc with ordering in $p_{\perp\mrm{ant}}^2$ is
  shown with medium solid lines, ordering in $m_\mrm{ant}^2$ with
  dashed lines and \Py{8}\ is shown with dotted lines.  The error bars
  limited by the horizontal lines indicate the statistical
  uncertainties, while the total uncertainties correspond to the full
  error bars.  The ratio plots show the deviation of the predictions
  from the data in units of the total uncertainty.}
\label{fig:massratiodata}
\end{figure}

In \FigsRef{fig:massratiodata} \tit{c)} and \tit{d)} we show the
asymmetry of the $\rho$ variable as a function of the dividing point
$\rho_0$. This asymmetry is sensitive to the relative number of
same-side versus opposite-side events, whose definitions were given in
\secRef{sec:observables}.  This asymmetry is seen to provide
discrimination between most of the shower models. The \Py8\ and \hpp
$\tilde q^2$ models yield predictions that lie within one standard
deviation of the data. However, the \hpp $q_\mrm{dip}^2$ model
predicts too small an asymmetry by about four standard deviations,
meaning that there are too few same-side compared to opposite-side
events. The two \vc models also predict too few same-side events, but
only at the level of around one standard deviation. In contrast, the
\hpp $p_{\perp\mrm{dip}}^2$ model predicts relatively too many same-side
events, at the level of two standard deviations.

\section{Summary and conclusion \label{sec:conclusions}}

We have presented measurements of distributions in
\epem\ annihilations at $\sqrt{s}=91.2~\mrm{GeV}$ that are sensitive
to QCD colour coherence, the ordering parameter in parton showers, and
to whether four-jet events arise from two separate $1\to2$ splittings
or from a $1\to3$ splitting. The data, corresponding to a sample of
about $397~000$ hadronic annihilation events, were collected with the
OPAL detector at LEP. The event selection criteria are defined in a
way to minimize the influence of non-perturbative (hadronization)
effects. We compared the data with six different models for the parton
shower, based on the \hpp, \Py8, and \vc Monte Carlo event generator
programs, which differ in the choice of the radiation function,
ordering variable, and recoil strategy. Each of the six models was
found to be in general agreement with the data. However, interesting
differences between the models and between some of the models and the
data were observed when asymmetries in the distributions were
examined.

Until now it was nearly impossible to distinguish between the
predictions of \Py{8} and \vc, or between the different variants of
\vc. Our study of the asymmetry of the ratio of squared jet masses,
shown in \FigRef{fig:massratiodata} \tit{d)}, shows that \vc predicts
somewhat too many opposite-side events (i.e., events with two $1\to2$
splittings) compared to same-side events (i.e., events with a $1\to3$
splitting), and that the data prefer \Py{8}. We find that the
different variants of \vc can be distinguished using the
central-to-towards (\FigRef{fig:angle14dataAS} \tit{b)}) and
central-to-away (\FigRef{fig:angle14dataAS} \tit{d)}) ratios in the
$\theta_{14}$ variable, which indicate that the \vc variant based on
antenna mass-squared evolution predicts somewhat too many wide-angle
emissions for the soft fourth jet, compared to collinear emissions.

To summarize the results of our study, we show the average p-value for
each model, calculated from the four values of the single observables,
in the bottom row of \TabRef{tab:chi2values}.  The variant of \hpp
with a $q^2_\mrm{dip}$-ordered dipole shower is found to
provide the least satisfactory description of the data. This model
does not contain coherence; it has intentionally been introduced to
confront it with coherent evolution. Thus our results emphasize the
importance of incorporating coherence into the description of the QCD
multijet process.  Since \hpp uses the cluster
\cite{Marchesini:1983bm} and \Py{8} and \vc use the Lund string
\cite{Andersson:1983ia,Sjostrand:1984ic} hadronization model, a direct
comparison of the predictions from the two groups of shower models is
somewhat ambiguous. It would be interesting to perform a comparison
based on use of the same hadronization model for all models.  However,
when comparing all shower models together, we find \Py{8} and \vc with
evolution in transverse momentum to give the best description of the
measurements presented here.

\subsection*{Acknowledgements}

We would like to express our gratitude to the members of the editorial
board (S. Bentvelssen, S. Bethke, J.W. Gary, K. Rabbertz) for their
careful review of the analysis and the draft resulting in this
publication.

NF would like to thank CERN for hospitality during the course of this
work.  This work was supported by the FP7 Marie Curie Initial Training
Network MCnetITN under contract number PITN-GA-2012-315877. SG and SP
acknowledge support from the Helmholtz Alliance ``Physics at the
Terascale''. SP acknoweldges support through a Marie Curie
Intra-European Fellowship under contract number PIEF-GA-2013-628739.

We particularly wish to thank the SL Division for the efficient
operation of the LEP accelerator at all energies and for their close
cooperation with our experimental group. In addition to the support
staff at our own institutions we are pleased to acknowledge the
Department of Energy, USA, \\
National Science Foundation, USA, \\
Particle Physics and Astronomy Research Council, UK, \\
Natural Sciences and Engineering Research Council, Canada, \\
Israel Science Foundation, administered by the Israel Academy of Science and Humanities, \\
Benoziyo Center for High Energy Physics, \\
Japanese Ministry of Education, Culture, Sports, Science and Technology (MEXT) and a grant
under the MEXT International Science Research Program, \\
Japanese Society for the Promotion of Science (JSPS), \\
German Israeli Bi-national Science Foundation (GIF), \\
Bundesministerium f{\"u}r Bildung und Forschung, Germany, \\
National Research Council of Canada, \\
Hungarian Foundation for Scientific Research, OTKA T-038240, and T-042864, \\
The NWO/NATO Fund for Scientific Research, the Netherlands.

\providecommand{\href}[2]{#2}\begingroup\raggedright

\endgroup

\end{document}